\documentclass[%
 aip,
 amsmath,amssymb,
 reprint,%
]{revtex4-1}

\usepackage{graphicx}
\usepackage{dcolumn}
\usepackage{bm}

\usepackage[utf8]{inputenc}
\usepackage[T1]{fontenc}
\usepackage{mathptmx}
\usepackage{etoolbox}
\usepackage{xcolor}
\usepackage{calrsfs}
\DeclareMathAlphabet{\pazocal}{OMS}{zplm}{m}{n}

\makeatletter
\def\@email#1#2{%
 \endgroup
 \patchcmd{\titleblock@produce}
  {\frontmatter@RRAPformat}
  {\frontmatter@RRAPformat{\produce@RRAP{*#1\href{mailto:#2}{#2}}}\frontmatter@RRAPformat}
  {}{}
}%
\makeatother
\begin{document}


\newcommand{\dudR}{\frac{\partial u}{\partial R}}
\newcommand{\dudphiR}{\frac{1}{R}\frac{\partial u}{\partial\phi}}
\newcommand{\dudphi}{\frac{\partial u}{\partial\phi}}
\newcommand{\dudZ}{\frac{\partial u}{\partial Z}}

\newcommand{\dudRt}{\frac{\partial u}{\partial  \tilde{R}}}
\newcommand{\dudxit}{\frac{\partial u}{\partial\tilde{\phi}}}
\newcommand{\dudZt}{\frac{\partial u}{\partial \tilde{Z}}}

\newcommand{\dBdR}{\frac{\partial B}{\partial R}}
\newcommand{\dBdphiR}{\frac{1}{R}\frac{\partial B}{\partial\phi}}
\newcommand{\dBdZ}{\frac{\partial B}{\partial Z}}
\newcommand{\dBdphi}{\frac{\partial B}{\partial \phi}}

\newcommand{\dBZdR}{\frac{\partial B_Z}{\partial R}}
\newcommand{\dBZdphi}{\frac{\partial B_Z}{\partial \phi}}
\newcommand{\dBZdZ}{\frac{\partial B_Z}{\partial Z}}

\newcommand{\dBRdR}{\frac{\partial B_R}{\partial R}}
\newcommand{\dBRdphi}{\frac{\partial B_R}{\partial \phi}}
\newcommand{\dBRdZ}{\frac{\partial B_R}{\partial Z}}

\newcommand{\dBphidR}{\frac{\partial B_{\phi}}{\partial R}}
\newcommand{\dBphidphi}{\frac{\partial B_{\phi}}{\partial \phi}}
\newcommand{\dBphidZ}{\frac{\partial B_{\phi}}{\partial Z}}

\newcommand{\dudRR}{\frac{\partial^2 u}{\partial R^2}}
\newcommand{\dudZZ}{\frac{\partial^2 u}{\partial Z^2}}
\newcommand{\dudphiphi}{\frac{1}{R^2}\frac{\partial^2 u}{\partial \phi^2}}
\newcommand{\dudRZ}{\frac{\partial^2 u}{\partial R\partial Z}}
\newcommand{\dudRphi}{\frac{1}{R}\frac{\partial^2 u}{\partial R\partial \phi}}
\newcommand{\dudphiZ}{\frac{1}{R}\frac{\partial^2 u}{\partial \phi\partial Z}}

\newcommand{\dudRtRt}{\frac{\partial^2 u}{\partial \widetilde{R}^2}}
\newcommand{\dudZtZt}{\frac{\partial^2 u}{\partial \widetilde{Z}^2}}
\newcommand{\dudRtZt}{\frac{\partial^2 u}{\partial \widetilde{R}\partial \widetilde{Z}}}
\newcommand{\dudxitxit}{\frac{\partial^2 u}{\partial \widetilde{\phi}^2}}
\newcommand{\dudRtxit}{\frac{\partial^2 u}{\partial \widetilde{R}\partial\widetilde{\phi}}}
\newcommand{\dudxitZt}{\frac{\partial^2 u}{\partial \widetilde{\phi}\partial\widetilde{Z}}}

\newcommand{\dudxi}{\frac{\partial u}{\partial \phi}}
\newcommand{\dudxixi}{\frac{\partial^2 u}{\partial \phi^2}}
\newcommand{\dudRxi}{\frac{\partial^2 u}{\partial R\partial\phi}}
\newcommand{\dudRxit}{\frac{\partial^2 u}{\partial R\partial\widetilde{\phi}}}
\newcommand{\dudxiZ}{\frac{\partial^2 u}{\partial \phi\partial Z}}
\newcommand{\dudxitZ}{\frac{\partial^2 u}{\partial \widetilde{\phi}\partial Z}}

\newcommand{\dphidR}{\frac{\partial \Phi}{\partial R}}
\newcommand{\dphidphiR}{\frac{1}{R}\frac{\partial \Phi}{\partial\phi}}
\newcommand{\dphidZ}{\frac{\partial \Phi}{\partial Z}}

\newcommand{\dphidRt}{\frac{\partial \Phi}{\partial  \tilde{R}}}
\newcommand{\dphidxit}{\frac{\partial \Phi}{\partial\tilde{\phi}}}
\newcommand{\dphidZt}{\frac{\partial \Phi}{\partial \tilde{Z}}}

\newcommand{\ls}{l_{\bot}}
\newcommand{\lp}{l_{\parallel}}

\newcommand{\BRB}{\frac{B_R}{B}}
\newcommand{\BphiB}{\frac{B_{\phi}}{B}}
\newcommand{\BZB}{\frac{B_Z}{B}}

\newcommand{\Lp}{L^{\parallel}}
\newcommand{\Ls}{L^{\bot}}

\newcommand{\umB}{\frac{1}{B}}
\newcommand{\umBtwo}{\frac{1}{B^2}}
\newcommand{\umR}{\frac{1}{R}}
\newcommand{\umRtwo}{\frac{1}{R^2}}
\newcommand{\umRO}{\frac{1}{R_0}}
\newcommand{\umROtwo}{\frac{1}{R_0^2}}

\newcommand{\Bphi}{B_{\phi}}

\newcommand{\eps}{\epsilon^{-1}}
\newcommand{\sigmaum}{\sigma^{-1}}
\newcommand{\sigmatwo}{\sigma^{-2}}
\newcommand{\larmor}{\rho_{s0}}

\newcommand{\eR}{\boldsymbol{e_R}}
\newcommand{\ephi}{\boldsymbol{e_{\phi}}}
\newcommand{\eZ}{\boldsymbol{e_Z}}

\newcommand{\dR}[1]{\frac{\partial}{\partial R}\left(#1\right)}
\newcommand{\dZ}[1]{\frac{\partial}{\partial Z}\left(#1\right)}
\newcommand{\dphiR}[1]{\frac{1}{R}\frac{\partial}{\partial \phi}\left(#1\right)}
\newcommand{\dphi}[1]{\frac{\partial}{\partial \phi}\left(#1\right)}

\newcommand{\dbetadR}{\frac{\partial\beta}{\partial R}}
\newcommand{\dbetadphi}{\frac{\partial\beta}{\partial \phi}}
\newcommand{\dbetadZ}{\frac{\partial\beta}{\partial Z}}

\newcommand{\dalphadR}{\frac{\partial\alpha}{\partial R}}
\newcommand{\dalphadphi}{\frac{\partial\alpha}{\partial \phi}}
\newcommand{\dalphadZ}{\frac{\partial\alpha}{\partial Z}}

\newcommand{\dgammadR}{\frac{\partial\gamma}{\partial R}}
\newcommand{\dgammadphi}{\frac{\partial\gamma}{\partial \phi}}
\newcommand{\dgammadZ}{\frac{\partial\gamma}{\partial Z}}

\newcommand{\rorhos}{\rho_*^{-1}}

\newcommand{\p}{\partial}

\newcommand{\pt}[1] {\frac{\partial #1}{\partial t}}

\newcommand{\nablapar}{\nabla_{\parallel}}
\newcommand{\nablaperp}{\nabla_{\bot}}
\newcommand{\vpare}{V_{\parallel e}}
\newcommand{\vpari}{V_{\parallel i}}

\newcommand{\jpar}{j_{\parallel}}

\newcommand{\rhos}{\rho_{s0}}

\newcommand{\kpar}{k_{\parallel}}
\newcommand{\kperp}{k_{\bot}}

\newcommand{\PLH}{{\mkern-2mu\times\mkern-2mu}}

\newcommand{\eo}{\boldsymbol{e_{1}}}
\newcommand{\et}{\boldsymbol{e_{2}}}
\newcommand{\etr}{\boldsymbol{e_{3}}}

\newcommand{\es}{\boldsymbol{e}_s}
\newcommand{\etheta}{\boldsymbol{e}_{\theta}}
\newcommand{\ethetas}{\boldsymbol{e}_{\theta^*}}

\newcommand{\bu}{\mathbf{b}}
\newcommand{\B}{\mathbf{B}}
\newcommand{\bunit}{\mathbf{b}}
\newcommand{\E}{\mathbf{E}}
\newcommand{\V}{\mathbf{V}}
\newcommand{\rp}{\mathbf{r}}
\newcommand{\vp}{\mathbf{v}}
\newcommand{\q}{\mathbf{q}}

\newcommand{\n}{\mathbf{\widehat{n}}}

\newcommand{\ex}{\mathbf{e_x}}
\newcommand{\ey}{\mathbf{e_y}}
\newcommand{\ez}{\mathbf{e_z}}

\newcommand{\detdz}{\frac{\partial \et}{\partial z}}
\newcommand{\detrdz}{\frac{\partial \etr}{\partial z}}

\newcommand{\gradx}{\mathbf{\nabla}x}
\newcommand{\grady}{\mathbf{\nabla}y}
\newcommand{\gradz}{\mathbf{\nabla}z}

\newcommand{\dfdx}{\frac{\partial f}{\partial x}}
\newcommand{\dfdy}{\frac{\partial f}{\partial y}}
\newcommand{\dfdz}{\frac{\partial f}{\partial z}}

\newcommand{\dphidx}{\frac{\partial \phi}{\partial x}}
\newcommand{\dphidy}{\frac{\partial \phi}{\partial y}}
\newcommand{\dphidz}{\frac{\partial \phi}{\partial z}}

\newcommand{\dtfdxt}{\frac{\partial^2f}{\partial x^2}}
\newcommand{\dtfdyt}{\frac{\partial^2f}{\partial y^2}}
\newcommand{\dtfdzt}{\frac{\partial^2f}{\partial z^2}}

\newcommand{\dtfdxdy}{\frac{\partial^2f}{\partial x\partial y}}
\newcommand{\dtfdxdz}{\frac{\partial^2f}{\partial x\partial z}}
\newcommand{\dtfdydz}{\frac{\partial^2f}{\partial y\partial z}}

\newcommand{\ddx}[1]{\frac{\partial}{\partial x}\left(#1\right)}
\newcommand{\ddy}[1]{\frac{\partial}{\partial y}\left(#1\right)}
\newcommand{\ddz}[1]{\frac{\partial}{\partial z}\left(#1\right)}

\newcommand{\dneqdx}{\frac{\partial n_0}{\partial x}}
\newcommand{\dphieqdx}{\frac{\partial \phi_0}{\partial x}}
\newcommand{\dTeqdx}{\frac{\partial T_{e0}}{\partial x}}

\newcommand{\EXB}{\textbf{E}\times\textbf{B}}

\newcommand{\Nfp}{N_{\text{fp}}}
\newcommand{\Raxis}{R_{\text{axis}}}
\newcommand{\Zaxis}{Z_{\text{axis}}}
\newcommand{\Mpol}{M_{\text{pol}}}
\newcommand{\Ntor}{N_{\text{tor}}}

\newcommand{\ehatx}{\widehat{\mathbf{e}}_x}
\newcommand{\ehaty}{\widehat{\mathbf{e}}_y}
\newcommand{\ehatz}{\widehat{\mathbf{e}}_z}

\newcommand{\ehatR}{\widehat{\mathbf{e}}_R}
\newcommand{\ehatphi}{\widehat{\mathbf{e}}_{\phi}}
\newcommand{\ehatZ}{\widehat{\mathbf{e}}_Z}

\newcommand{\dRdtheta}{\frac{\partial R}{\partial\theta}}
\newcommand{\dRdphi}{\frac{\partial R}{\partial\phi}}

\newcommand{\dZdtheta}{\frac{\partial Z}{\partial\theta}}
\newcommand{\dZdphi}{\frac{\partial Z}{\partial\phi}}

\newcommand{\thetas}{\theta^*}

\newcommand{\Baxis}{B_{\text{axis}}}

\newcommand{\VE}{\mathbf{V}_{\text{E}\times\text{B}}}
\newcommand{\Vde}{\mathbf{V}_{\text{de}}}
\newcommand{\Vdi}{\mathbf{V}_{\text{di}}}
\newcommand{\Vpol}{\mathbf{V}_{\text{pol}}}

\newcommand{\blue}[1]{{\color{blue}#1}}
\newcommand{\red}[1]{{\color{red}#1}}
\newcommand{\green}[1]{{\color{green}#1}}
\newcommand{\black}[1]{{\color{black}#1}}

\preprint{AIP/123-QED}

\title[Magnetic shear effects on ballooning
turbulence in the boundary of fusion devices]{Magnetic shear effects on ballooning
turbulence in the boundary of fusion devices}
\author{Z. Tecchiolli}
\email{zeno.tecchiolli@epfl.ch}
\affiliation{ 
Ecole Polytechnique Fédérale de Lausanne (EPFL), Swiss Plasma Center (SPC), CH-1015 Lausanne, Switzerland
}%

\author{A. J. Coelho}
\affiliation{Gauss Fusion GmbH, Garching bei München, Germany}
\author{J. Loizu}
\affiliation{ 
Ecole Polytechnique Fédérale de Lausanne (EPFL), Swiss Plasma Center (SPC), CH-1015 Lausanne, Switzerland
}%
\author{B. De Lucca}
\affiliation{ 
Ecole Polytechnique Fédérale de Lausanne (EPFL), Swiss Plasma Center (SPC), CH-1015 Lausanne, Switzerland
}%
\author{P. Ricci}
\affiliation{ 
Ecole Polytechnique Fédérale de Lausanne (EPFL), Swiss Plasma Center (SPC), CH-1015 Lausanne, Switzerland
}%

\date{\today}

\begin{abstract}
The effect of magnetic shear on ballooning-driven plasma edge turbulence is studied through nonlinear simulations complemented by linear numerical and analytical investigations. Nonlinear, 3D, global, flux-driven simulations using the GBS code show that the scale separation between radial, $x$, and poloidal, $y$, size of turbulent eddies, \(k_x \ll k_y\), considered by \citet{paolo_rogers_brunner_2008} and extensively used to predict pressure gradient lengths, SOL width, particle and heat fluxes, is observed with high magnetic shear. In contrast, for low magnetic shear, \(k_x \sim k_y\) is observed, with fluctuation properties resembling those shown by recent low-shear stellarator simulations reported in Coehlo, Loizu, Ricci, and Tecchiolli\cite{GBS_stellarators}. Global linear investigations of the ballooning mode qualitatively captures the transition in mode structure with varying magnetic shear, showing that $k_x \ll k_y$ is achieved with sufficiently strong poloidal mode coupling enhanced by increasing magnetic shear, resistivity, toroidal mode number, and equilibrium gradient scale length. This is confirmed by an analytical study considering a dominant poloidal mode and its sidebands, which highlights that the poloidal mode structure is determined by curvature and $k_\parallel$ effects.
\end{abstract}

\maketitle
\section{Introduction}
\label{sec:Introduction}

Turbulence in the boundary of tokamaks shows the presence of small-scale eddies originating inside the last closed flux surface (LCFS), that may develop into filaments detaching and propagating into the far scrape-off layer (SOL) \citep{endler1999turbulent, garcia2009blob}. The size of these eddies in the radial and binormal ($\sim$ poloidal) direction is of crucial importance for determining the plasma dynamics in the boundary region \citep{zeiler1996three, rogers1998phase,  maurizio_turbulent_regimes}.

The eddy size combined with quasi-linear theories \citep{zeiler} allows for the prediction of fundamental quantities such as the equilibrium pressure gradient
length and the SOL width \citep{ricci2010turbulence}. Indeed, the evaluation of the eddy size is a key element to estimate the particle and heat turbulent fluxes as it affects the turbulent regimes in the boundary \citep{ricci2010turbulence, annamaria_turbulent_regimes, halpern2013ideal, maurizio_turbulent_regimes} and the plasma-wall interaction \cite{loarte2007power}. Ultimately, operational limits and turbulent regime transitions such as the density and $\beta$ operational limits, and of the L-H transition \citep{ricci_rogers_2013,giacomin2022first, giacomin2022turbulent} depend on the shape of the turbulent eddies. Scaling laws for the equilibrium lengths and confinement times based on an estimate of the eddy size were successfully validated against multi-machine databases \citep{giacomin2022first, ricci2023theoretical, lim2023effect, lim2024predictive}. 

Previous studies
\citep{paolo_rogers_brunner_2008,rogers_dorland_2005,ricci_rogers_2013} have investigated the radial extension of the turbulent eddies in ballooning mode (BM) \citep{paolo_rogers_brunner_2008} and drift waves (DWs) \citep{rogers_dorland_2005} driven turbulence. For BM and DWs, the relationship \(k_x \sim \sqrt{k_y/L_p}\) is found analytically, where \(k_x\) and \(k_y\) denote radial and binormal wavenumbers, and \(L_p\) is the time-averaged pressure radial scale length. For Kelvin-Helmholtz turbulence \citep{rogers_dorland_2005,ricci_rogers_2013}, expected to be significant only in the presence of strong \(E \times B\) shear flow \citep{maurizio_turbulent_regimes}, the scaling \(k_x \sim 0.6k_y\) is observed. 

To simplify the study of eddy properties, mode locality assumptions are often made in linear studies of edge instabilities. In the case of ballooning modes (BMs), a key scaling relation, $
k_x \sim \sqrt{k_y/L_p}$,
is derived by assuming strong poloidal localization, which holds when $ k_y L_p \gg 1 $ and a single dominant poloidal mode is considered \citep{rogers_dorland_2005,paolo_rogers_brunner_2008}.
The relation \( k_x \sim \sqrt{k_y/L_p} \) has been widely used to simplify the eddies analysis, as it ensures \( k_x \ll k_y \), allowing radial derivatives of the mode to be neglected. In fact, linear studies often adopt a flux-tube approach \citep{zeiler,rogers_dorland_2005,annamaria_linear_theory,halpern2013ideal}, which assumes $1/L_p \ll k_x, k_y$.

In contrast to tokamak simulations, recent, global, flux-driven, two-fluid, plasma turbulence simulations of an island-diverted stellarator \citep{letter_stellarators} show a BM dominated regime with $k_x \sim k_y$ eddies and fluctuation scales comparable to equilibrium gradients, $k_y \sim 1/L_p$. Fluctuations with these properties are also found experimentally in the TJ-K stellarator \citep{krause2002torsatron}, a feature successfully reproduced by global turbulent simulations \citep{TJK_mine}. The island diverted configuration and TJ-K both have a small global magnetic shear, $s \equiv (r/q) d q /d r \ll 1$ with $r$ being the radial direction and $q(r)$ the safety factor, while tokamak configurations are characterised by high magnetic shear, $s \sim 2-4$. These recent results motivate a detailed investigation of the size of turbulent eddies as a function of the magnetic shear.

This paper presents low magnetic-shear and a high magnetic-shear simulations of plasma in the edge region of a simple tokamak with circular flux surfaces. 
Given the high collisionality in the boundary region and being interested in turbulence time scales, the plasma dynamics is described by the drift-reduced Braginskii equations \citep{zeiler} evolved by GBS, a three-dimensional, global, two-fluid, flux-driven code \citep{coelho2024global}. Simulations parameters are chosen to ensure that turbulence is driven by BM, by tuning the ratio between collisionality and mass-ratio \citep{annamaria_turbulent_regimes}. Results show that, as magnetic shear decreases, the dominant mode wavenumber \(k_y\) decreases, and \(k_x \sim \sqrt{k_y/L_p}\) no longer holds. Indeed, the properties of the fluctuations recall those of recent low-shear stellarator simulations, where \(k_x \sim k_y\) \citep{letter_stellarators,TJK_mine}. These simulations show that, at low-shear, radial and poloidal fluctuations extend over a region comparable to \( L_p \), breaking the validity of the flux-tube approximation. Therefore, a global theory is needed to accurately describe the spatial structure of BMs in general magnetic shear conditions, without assuming a fixed relation between \( L_p \) and fluctuation size.

We introduce a novel global 3D linear theory demonstrating that, in low-shear regimes, radial and poloidal fluctuations are comparable in size, while $k_x \sim \sqrt{k_y/L_p}$ is recovered in high-shear regimes. The analytical theory supports our numerical results, pointing out that the turbulent mode poloidal structure results from the interplay between magnetic curvature, inducing mode coupling, and parallel gradient effects.

The paper is organized as follows. Section \ref{Sec:GBS_Code} describes the drift-reduced Braginskii equations used for simulating plasma turbulence in the boundary of fusion devices and the GBS code we use to evolve them. Section \ref{Sec:EquilibriumField} outlines the tokamak configurations with circular flux surfaces and varying magnetic shear considered for our study, and discusses nonlinear results of ballooning-driven turbulence. Section \ref{Sec:Linear} details the extended linear ballooning theory in two dimensions, comparing numerical results with a side-band analytical theory. Finally, Section \ref{Sec:Discussion} provides a discussion of our findings.

\section{The drift-reduced Braginskii equations  }\label{Sec:GBS_Code}

We consider a magnetized plasma modelled by the drift-reduced Braginskii equations, valid at high collisionality under the assumption of perpendicular scale lengths of the dominant modes larger than the ion Larmor radius. This regime often characterizes the plasma boundary of magnetic fusion devices, as well as the core of low-temperature devices such as the TORPEX basic plasma physics experiment \citep{torpex_galassi} or the TJ-K stellarator \citep{TJK_mine}. This set of equations is evolved by the GBS code \citep{paolo_GBS,GBS_cite_Maurizio,halpern}, a three-dimensional, global, two-fluid, flux-driven code that solves the drift-reduced Braginskii equations \citep{zeiler}. GBS evolves all quantities in time, without separation between equilibrium and fluctuating components. In this work, the electrostatic limit of the model is taken, the Boussinesq approximation is employed \citep{paolo_GBS} and the gyroviscous terms, as well as the plasma coupling to the neutral dynamics, are neglected, although these are implemented in the most recent version of the GBS code for tokamak simulations \citep{GBS_cite_Maurizio}. Within these approximations, the set of equations considered in this work is
\begin{align}
    &\pt{N} = -\frac{\rorhos}{B}\left[\Phi,N\right] - \nablapar(N \vpare) + \frac{2}{B}\left[C(p_e)-N C(\Phi)\right]  \notag \\  & \hspace{1cm} + D_N\nabla_{\perp}^2N   +D_N^{\parallel} \nablapar^2N +  S_n  \label{eq:GBSn} \\
    & \pt{T_e} = -\frac{\rorhos}{B}\left[\Phi,T_e\right] - \vpare\nablapar T_e + \frac{4T_e}{3B}\Big[\frac{C(p_e)}{N}+\frac{5}{2}C(T_e) \notag \\ & \hspace{1cm}-C(\Phi)\Big]  + \frac{2T_e}{3 N}\left[0.71\nablapar j_{\parallel}-N\nablapar\vpare\right] \notag \\ & \hspace{1cm}  + D_{T_e}\nablaperp^2T_e+\chi_{\parallel e}\nablapar^2T_e + S_{T_e} \label{eq:GBSTe}\\
    &\pt{T_i} = -\frac{\rorhos}{B}\left[\Phi,T_i\right] - \vpari\nablapar T_i + \frac{4T_i}{3B}\Bigg[\frac{C(p_e)}{N}-\frac{5}{2}\tau C(T_i) \notag \\ & \hspace{1cm}-C(\Phi)\Bigg] + \frac{2T_i}{3 N}\left[\nablapar j_{\parallel} - N\nablapar\vpari\right] + D_{T_i}\nablaperp^2T_i \notag \\ & \hspace{1cm} + \chi_{\parallel i}\nablapar^2T_i +S_{T_i} \label{eq:GBSTi}
\end{align}
 \begin{align}
    & \pt{\vpare} = -\frac{\rorhos}{B}\left[\Phi,\vpare\right]-\vpare\nablapar\vpare + \frac{m_i}{m_e}\Bigg[\nu j_{\parallel}+\nablapar\Phi \notag\\ & \hspace{1cm} -\frac{\nablapar p_e}{N} - 0.71\nablapar T_e\Bigg] + \eta_{0e}\nablapar^2\vpare + D_{\vpare}\nablaperp^2\vpare \label{eq:GBSVe}\\
    & \pt{\vpari} = -\frac{\rorhos}{B}\left[\Phi,\vpari\right]-\vpari\nablapar\vpari - \frac{1}{N}\nablapar(p_e+\tau p_i)\notag \\ &\hspace{1cm} +\eta_{0i}\nablapar^2\vpari   +D_{\vpari}\nablaperp^2\vpari \label{eq:GBSVi}\\
    &\pt{\omega} = -\frac{\rorhos}{B}\left[\Phi,\omega\right] -\vpari\nablapar\omega + \frac{B^2}{N}\nablapar j_{\parallel} + \frac{2B}{N}C(p_e+\tau p_i)\notag \\ &\hspace{1cm} + D_{\omega}\nablaperp^2\omega   + D_{\omega}^{\parallel}\nablapar^2\omega \label{eq:GBSomega}
\end{align}
which are closed by
\begin{equation}
    \nablaperp^2\Phi = \omega - \tau\nablaperp^2T_i.
    \label{eq:GBS_phi}
\end{equation}

In Eqs. ~(\ref{eq:GBSn}-\ref{eq:GBS_phi}) and in the rest of the paper, the density $N$, the electron temperature $T_e$, the ion temperature $T_i$ and the norm of the magnetic field $B$ are normalized to the reference values $N_0$, $T_{e0}$, $T_{i0}$ and $B_0$, respectively; electron parallel velocity $\vpare$ and ion parallel velocity $\vpari$ are both normalized to the sound speed $c_{s0}=\sqrt{T_{e0}/m_i}$; vorticity $\omega$ and electrostatic potential $\Phi$ are respectively normalized to $T_{e0}/(e\rhos^2)$ and $T_{e0}/e$; time is normalized to $R_0/c_{s0}$, where $R_0$ is a reference macroscopic scale length, typically the major radius of the machine; perpendicular and parallel lengths are normalized to the ion sound Larmor radius, $\rhos=\sqrt{T_{e0}m_i}/(eB_0)$, and $R_0$, respectively. The normalized parallel current density is $j_{\parallel}=N(\vpari-\vpare)$.

The dimensionless parameters appearing in Eqs.~(\ref{eq:GBSn}-\ref{eq:GBS_phi}) are the normalized ion sound Larmor radius $\rho_*=\rhos/R_0$, the normalized electron and ion parallel heat diffusivities, $\chi_{\parallel e}$ and $\chi_{\parallel i}$, considered constant here, the ion to electron temperature ratio $\tau=T_{i0}/T_{e0}$, the normalized electron and ion viscosities, $\eta_{0e}$ and $\eta_{0i}$, which we assume to have constant values, and the normalized Spitzer resistivity, $\nu=\nu_0T_e^{-3/2}$, with
\begin{equation}
    \nu_0=\frac{4\sqrt{2\pi}}{5.88}\frac{e^4}{(4\pi\varepsilon_0)^2}\frac{\sqrt{m_e}R_0 N_0\lambda}{m_ic_{s0}T_{e0}^{3/2}},
\end{equation}
\noindent
where $\lambda$ denotes the Coulomb logarithm \citep{maurizio_turbulent_regimes}. Small numerical diffusion terms such as $D_N\nablaperp^2 N$ and $D_N^{\parallel}\nablapar^2 N$ (and similar for the other fields) are introduced to improve the numerical stability of the simulations. The simulation results show that they have a negligible effect the plasma dynamics since they lead to significantly lower perpendicular transport than the turbulent phenomena.
The terms $S_N$, $S_{T_e}$ and $S_{T_i}$ denote the density, electron temperature and ion temperature sources, respectively. Magnetic pre-sheath boundary conditions \citet{joaquim_BCs, annamaria_BCs} are applied to all quantities on the domain boundaries where the plasma is exhausted (the top and bottom domain boundaries in this work). A summary of the applied boundary conditions is given in Table ~\ref{Tab:BCs}.

\begin{table}
\centering
\begin{tabular}{lcc}
\hline
 & Top/Bottom walls & Inner/Outer walls \\ \hline
 $\vpare$ & $\vpare=\pm \sqrt{T_e}\exp{(\Lambda-\Phi/T_e)}$ & $\partial_s \vpare=0$ \\ \hline
 $\vpari$ & $\vpari=\pm \sqrt{T_e}F_T$ & $\partial_s \vpari=0$ \\ \hline
 $\omega$ & $\omega=- \left [ \frac{(\partial_s v_{\parallel i})^2}{F_T^2}\pm \frac{\sqrt{T_e}}{F_T}\partial_s^2v_{\parallel i} \right ]$ & $\omega=0$ \\ \hline
 $n$ & $\partial_s n=\mp \frac{n}{\sqrt{T_e}F_T}$ & $\partial_s n=0$ \\ \hline
 $\Phi$ & $\partial_s\Phi=\pm\frac{\sqrt{T_e}}{F_T}\partial_s v_{\parallel i}$ / $\Phi=\Lambda T_e$ & $\Phi=\Lambda T_e$ \\ \hline
 $T_e,T_i$ & $\partial_s T_e=\partial_s T_i=0$ & $\partial_s T_e=\partial_s T_i=0$ \\ \hline
\end{tabular}
\caption{Boundary conditions for the simulations presented in this paper. The derivative $\partial_s = \mathbf{s}\cdot \boldsymbol{\nabla}$ is along the direction normal to the surface.}
\label{Tab:BCs}
\end{table}

Considering a generic scalar quantity $f$, the normalized geometrical operators appearing in Eqs.~(\ref{eq:GBSn}-\ref{eq:GBS_phi}) are the parallel gradient $\nabla_{\parallel}f = \boldsymbol{b}\cdot\boldsymbol{\nabla}f$, the Poisson brackets, $[\Phi,f]=\boldsymbol{b}\cdot\left[\boldsymbol{\nabla}\Phi\times\boldsymbol{\nabla} f\right]$, the curvature operator, $C(f) = (B/2)\left[\boldsymbol{\nabla}\times(\boldsymbol{b}/B)\right]\cdot\boldsymbol{\nabla}f$, the parallel Laplacian, $ \nabla_{\parallel}^2f = \boldsymbol{b}\cdot\boldsymbol{\nabla}(\boldsymbol{b}\cdot\boldsymbol{\nabla}f)$, and the perpendicular Laplacian $\nabla_{\bot}^2 f = \boldsymbol{\nabla}\cdot\left[(\boldsymbol{b}\times\boldsymbol{\nabla}f)\times\boldsymbol{b}\right]$. By assuming $B_p/B\ll1$, where $B_p$ is the poloidal magnetic field, and $l_{\bot}/l_{\parallel}\ll1$, where $l_{\bot}$ and $l_{\parallel}$ are the perpendicular and parallel turbulence length scales, and only retaining the leading order terms, the geometrical operators in an axisymmetric field in cylindrical coordinates $(R,\varphi,Z)$ become \citep{GBS_cite_Maurizio}:

\begin{align}
    \nabla_{\parallel}f & = \BRB\rorhos\frac{\partial f}{\partial R}+\texttt{sign}\left(\frac{B_\varphi}{B}\right)\frac{R_0}{R}\frac{\partial f}{\partial \varphi}+\BZB\rorhos \frac{\partial f}{\partial Z}
    \label{eq:parallel_gradient}\\
    [\Phi,f] & =\texttt{sign}\left(\frac{B_\varphi}{B}\right)\Bigg[\dphidZ\frac{\partial f}{\partial R} - \texttt{sign}\left(\BphiB\right)\dphidR\frac{\partial f}{\partial Z}\Bigg]
    \label{eq:poisson_brackets}\\
    C(f) & =  \texttt{sign}\left(\frac{B_\varphi}{B}\right)\frac{\partial f}{\partial Z}
    \label{eq:curvature_operator} \\
    \nabla^2_{\bot}f & = \frac{\partial^2 f}{\partial R^2} + \frac{\partial^2 f}{\partial Z^2}\label{eq:perpendicular_laplacian}
\end{align}

The parallel gradient is normalised to $1/R_0$, the perpendicular Laplacian to $1/R_0^2$, the Poisson brackets to $1/\rho_{s0}^2$, and the curvature to $1/R_0\rho_{s0}$. The GBS simulation domain is a torus of radius $R_0$ with a rectangular cross-section of size $L_R\times L_Z$. 

The physical model in Eqs.~(\ref{eq:GBSn}-\ref{eq:GBS_phi}) is discretized using a regular cylindrical grid. Equations (\ref{eq:GBSn}-\ref{eq:GBSomega}) are advanced in time using an explicit Runge-Kutta fourth-order scheme, while spatial derivatives are computed with a fourth-order finite difference scheme.

\section{Simulations setup and nonlinear results} \label{Sec:EquilibriumField}

\begin{figure}
 \centering
 \includegraphics[scale=0.35]{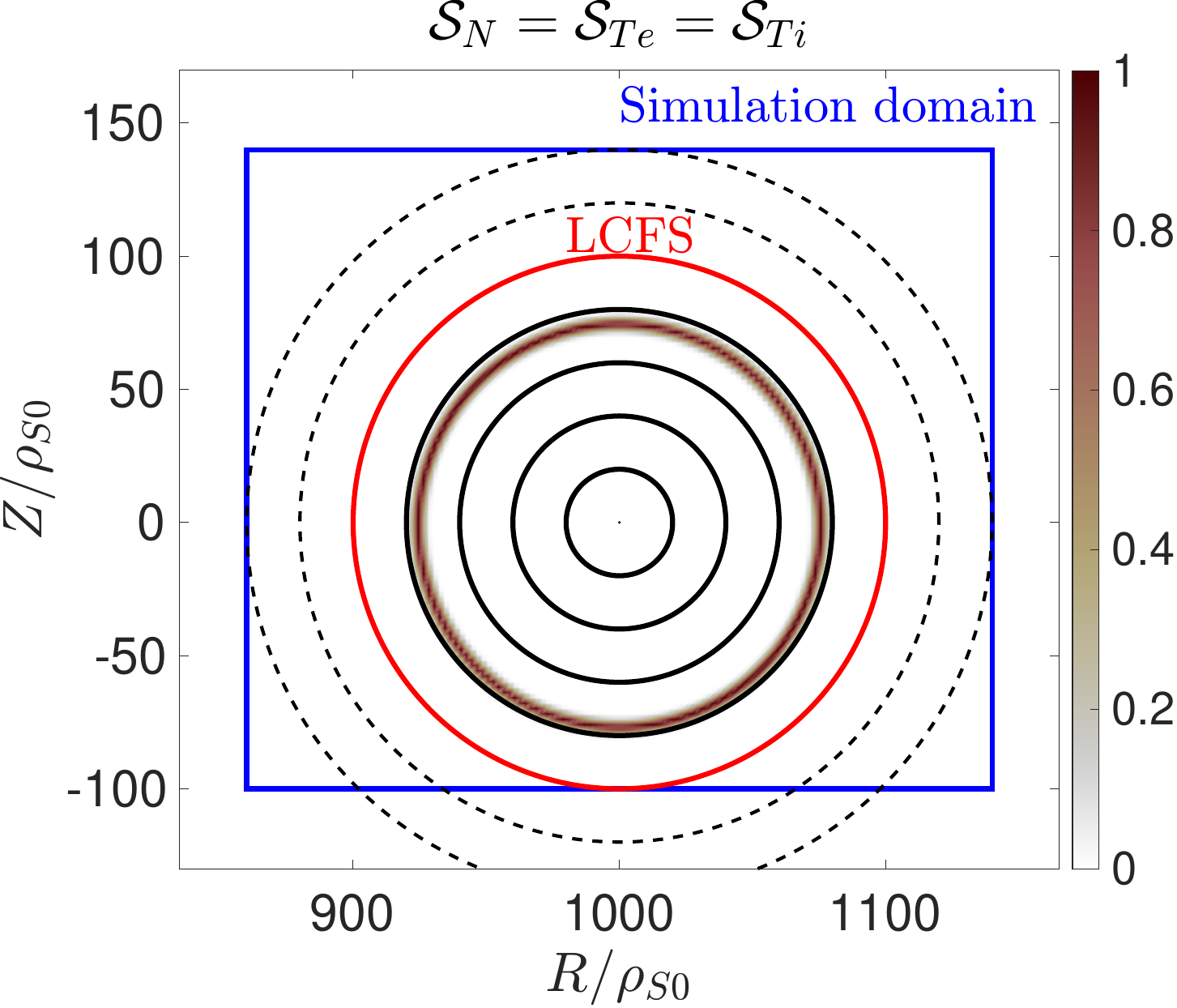}
 \caption{Poincar\'e plot of the tokamak magnetic field Eqs. (\ref{eq:magnetic_field_circular_tokamak1}-\ref{eq:magnetic_field_circular_tokamak2}). The GBS simulation domain is shown in blue. Closed flux surfaces are represented by continuous lines and open flux surfaces by dashed black lines. These are separated by the LCFS in red. The colorscale is for density, electron temperature and ion temperature sources.}
 \label{fig:tokamak_circular_flux_surfaces}
\end{figure}
\begin{figure*}
 \centering
\includegraphics[scale=0.5]{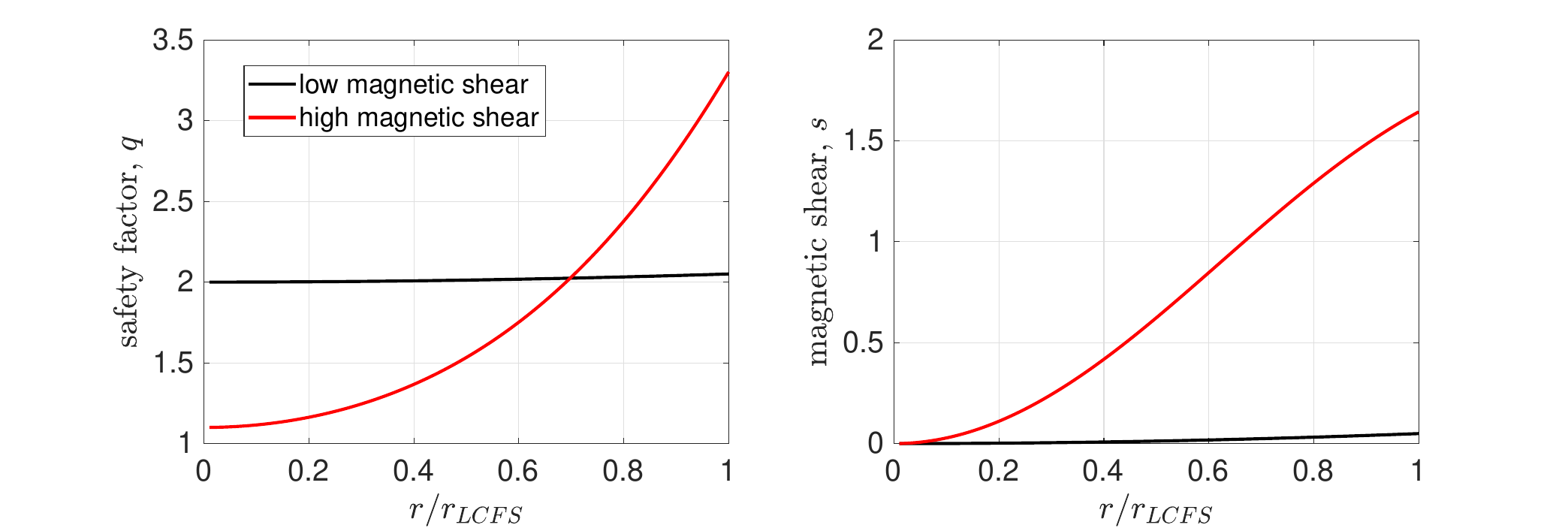}
 \caption{Safety factor and magnetic shear of the two circular tokamak configurations. In the low magnetic-shear case $s_a=0.05$, while in the high shear case $s_a=1.6$.}
 \label{fig:q_and_shear_tokamak_circular}
\end{figure*}

In the present work, our simulations consider an axisymmetric vacuum magnetic field with circular flux surfaces, assuming a toroidal current density, $J$, centred around $(R_0,Z_0)$, and described by a Gaussian profile, $J \sim e^{- r^2/w ^2}$, with $r$ the poloidal distance from $(R_0, Z_0)$ and $w$ a constant. This results is a magnetic field whose components in cylindrical coordinates are given by
\begin{align}
    &B_R = B_0\varsigma\frac{Z-Z_0}{(R-R_0)^2+(Z-Z_0)^2}  h(R,Z), \label{eq:magnetic_field_circular_tokamak1} \\ 
    &B_{\varphi} = B_0R_0/R, \\ \label{eq:magnetic_field_circular_tokamak2}
    &B_Z =-B_0\varsigma\frac{R-R_0}{(R-R_0)^2+(Z-Z_0)^2}  h(R,Z),
\end{align}
with $h(R,Z) = 1-\exp\left\{-\left[(R-R_0)^2+(Z-Z_0)^2/w^2\right]\right\}$, $B_0$ the magnetic field on axis, and $\varsigma$ is constant representing the typical length scale for the poloidal field. The magnetic field in Eqs. (\ref{eq:magnetic_field_circular_tokamak1}-\ref{eq:magnetic_field_circular_tokamak2}) generates circular flux surfaces (as shown  Fig.~\ref{fig:tokamak_circular_flux_surfaces}) and corresponds to the one considered in the $s-\alpha$ geometry with $\alpha = 0$ \citep{lapillonne2009clarifications}, since we do not include any Shafranov shift and electromagnetic effects. Previous two-fluid turbulence simulations in stellarators\cite{TJK_mine, coelho2024global} also considered electrostatic turbulence on a fixed vacuum field. This configuration was previously employed for studies of the SOL instabilities and turbulent regimes in limited plasmas \citep{rogers1998phase,annamaria_linear_theory, annamaria_turbulent_regimes}, showing no instability threshold for the resistive BM.

The parameters $\varphi$ and $w$ are adjusted to fix the values of the safety factor $q$ at the magnetic axis and at the LCFS ($r=a$). In fact, since $B^{\theta}=B_R\sin\theta/r-B_Z\cos\theta/r$ and $B^{\varphi}=B_{\varphi}/R$, and considering that flux surfaces are circular, the safety factor follows as
\begin{align} \label{eq:qfactnonlin}
    q(r) = \frac{1}{2\pi}\int_0^{2\pi}\frac{B^{\varphi}}{B^{\theta}}d\theta = \frac{1}{R_0\varphi}\frac{r^2}{1-\exp(- r^2/w^2)}
\end{align}
in the infinite aspect-ratio limit,

 where $(r, \varphi, \theta)$ are the right-handed toroidal coordinates. Thus, the safety factor on axis is $q_0=w^2/(R_0\varphi)$, which allows us to write $w^2=R_0\varphi q_0$. The value of $q$ at the LCFS, $q(r=a)=q_a$, allows us to obtain the parameter $\varphi$ by solving Eq. (\ref{eq:qfactnonlin}) in the large aspect ratio. 
The magnetic shear, $s=(r/q)dq/dr$, yields
\begin{equation}\label{eq:shearnonlin}
    s(r) = 2 + \frac{2r^2}{R_0\varphi q_0\left[1-\exp\left(\frac{r^2}{R_0\varphi q_0}\right)\right]}.
\end{equation}
We note that $s(r=0)=0$ and $s(r\rightarrow\infty)=2$, i.e. the magnetic shear is bounded by $2$ in the equilibria considered here.

For our investigations, we consider two configurations: a low magnetic-shear with $q_0=2$ and $q_a=2.05$, and a high magnetic-shear with $q_0=1.1$ and $q_a=3.3$. By setting $R_0=1000\rhos$ and $a=100\rhos$, this requires $\varphi=100\rhos$ and $w=447\rhos$ in the low magnetic-shear case, while $\varphi=3.22\rhos$ and $w=59.3\rhos$ in the high magnetic-shear case. The safety factor and magnetic shear profiles for the two configurations are shown in Fig.~\ref{fig:q_and_shear_tokamak_circular}. The values of magnetic shear at the LCFS in the two configurations are $s_a=0.05$ and $s_a=1.6$. The high shear value is in line with typical edge shear values for tokamak discharges, $s \approx 1.0-4.0$, such as in the case of TCV \citep{piras2010snowflake, oliveira2022validation, reimerdes2022overview}. While fusion-relevant tokamaks operate with large magnetic shear at the edge, basic plasma studies can be performed in tokamaks with low-shear profiles. This is the case of the Madison Symmetric Torus (MST) operated in a tokamak configuration \cite{munaretto2020generation}, able to nondisruptively operate\cite{hurst2022self} with $q$-value
at the edge between $0.8 - 2$ while $q$ on the axis is $1$ with low normalised $\beta < 1\%$. Nevertheless, owing to its thick, stabilizing, conductive wall, MST reached tokamak plasmas with density up to $10$ times the Greenwald limit\cite{hurst2024tokamak} with low $q$ at the edge.

The GBS simulation parameters considered here are $\rorhos=1000$, $\nu_0=0.1$, $m_i/m_e=200$, $\tau=1$, $\chi_{\parallel e,i}=\eta_{0 e,i}=1.0$, $D_N = D_{Te} = D_{Ti} = D_{V \parallel e} = D_{V\parallel i} = D_{\omega} = 10$, $D_{N}^{\parallel}=D_{\omega}^{\parallel}=1$, a grid resolution of $\Delta R=\Delta Z=2\rhos$, $\Delta\phi=2\pi/80$, and a time-step of the order of $2.0\times 10^{-5}R_0/c_{s0}$. The density and temperature sources are equal and correspond to a radially localized Gaussian around a closed flux surface near the LCFS, as shown in Fig.~\ref{fig:tokamak_circular_flux_surfaces}. Starting from an initial state with constant profiles, a quasi-steady state is reached after a transient where sources, parallel and perpendicular transport, and losses at the walls balance each other. To ensure statistically meaningful results, we time average quantities within a window of $20$ time units during the steady state. Time average quantities are indicated with $\langle f \rangle_t$, and time fluctuations are defined as $\Tilde{f}\equiv f - \langle f \rangle_t$.

\begin{figure}[h]
 \centering
 \includegraphics[scale=0.50]{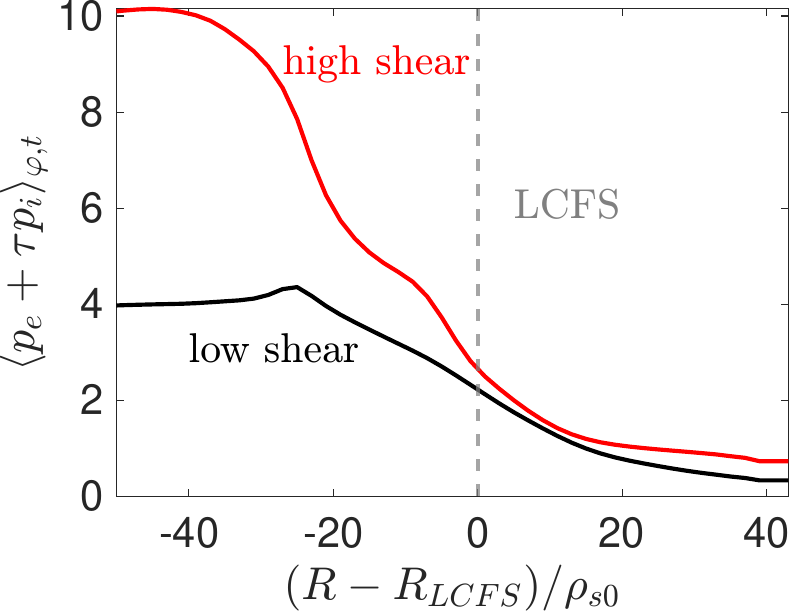}
 \caption{Toroidally and time-averaged pressure profiles $\langle p_e + \tau p_i\rangle_{\varphi, t}$ close to the LCFS at $Z = 0$ for the high-shear (red) and low-shear (black) cases.}
 \label{fig:PressurePlot}
 \end{figure}
\begin{figure*}
 \centering
 \includegraphics[scale=0.4]{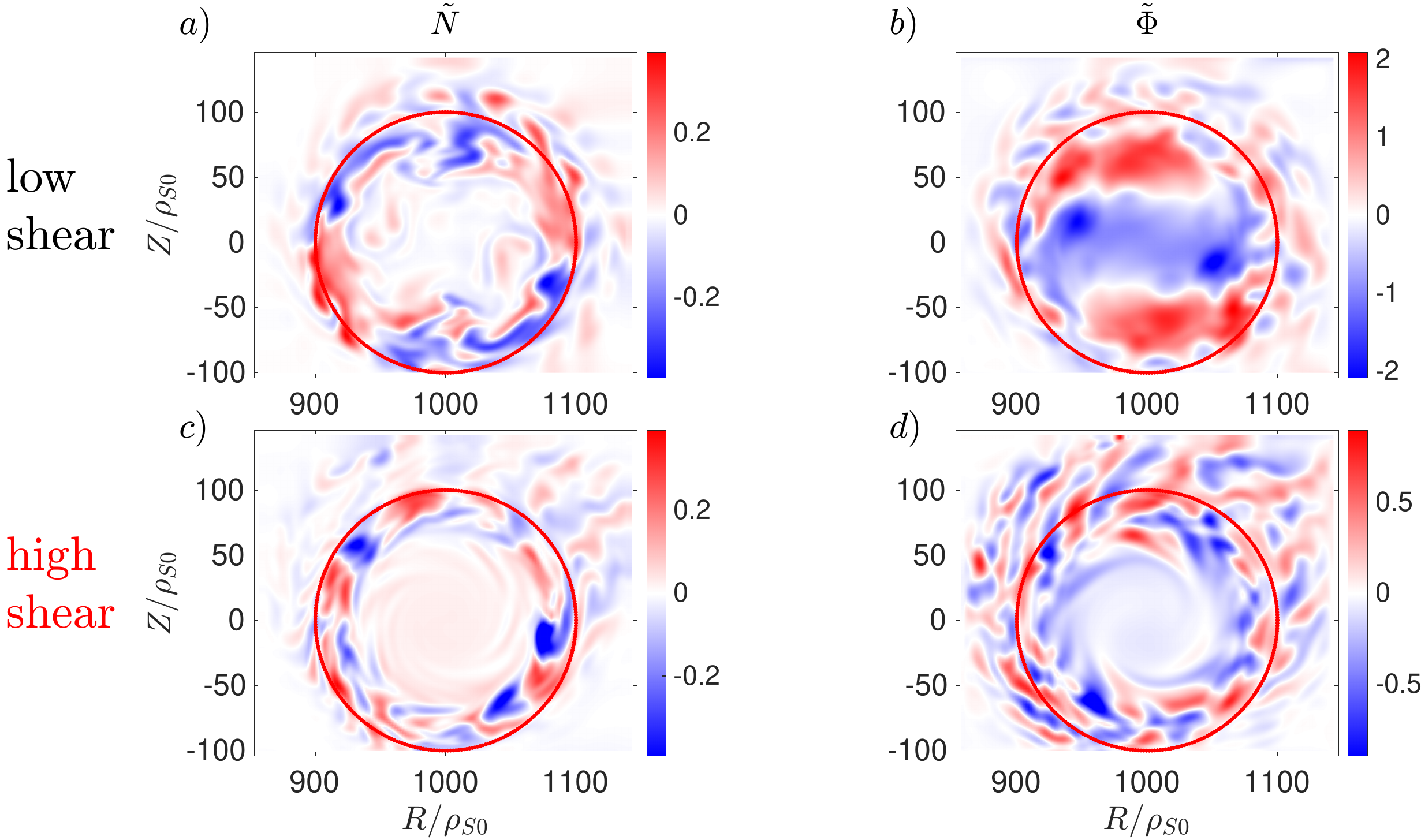}
 \caption{Typical density and potential fluctuations of low ($s_a=0.05$) and high ($s_a=1.6$) magnetic shear tokamak configurations.}
 \label{fig:fluctuations_tokamak_circular}
\end{figure*}
\begin{figure}
 \centering
 \includegraphics[scale=0.30]{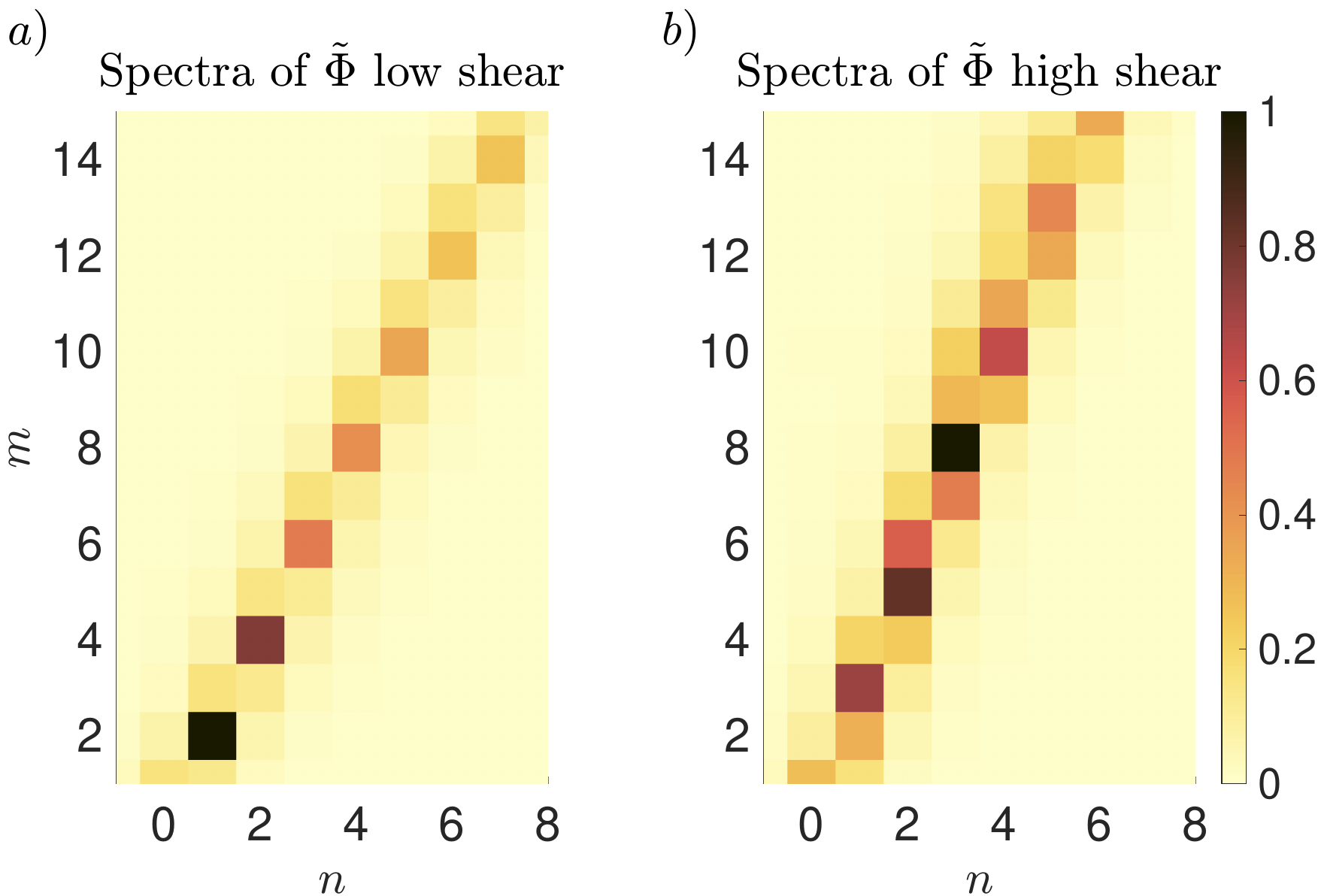}
 \caption{Time-averaged Fourier spectrum of potential fluctuations on the LCFS in the two tokamak configurations with low, $s_a=0.05$  in $(a)$, and high, $s_a=1.6$ in $(b)$, magnetic shear.}
 \label{fig:fourier_spectrum_tokamak}
 \end{figure}

As shown in Fig. \ref{fig:PressurePlot}, the high-shear configuration shows higher peak pressure values in the core region than the low-shear configuration, leading to high energy confinement time $\tau_E = \int \langle p_e + \tau p_i \rangle_{t} dV/ \int S_p dV$, since the power source $S_p$ is the same in the two configurations. On the other hand, $L_p \sim - p/\nabla p$ are comparable close to the LCFS, with $L_p = 20.3$ in the low-shear case and $L_p = 24.1$ in the high-shear case.

A typical snapshot of density and potential fluctuations, $\Tilde{N}$ and $\tilde{\Phi}$, of the low and high shear tokamak configurations are shown in Fig. \ref{fig:fluctuations_tokamak_circular}. The low magnetic-shear simulation shows larger structures, in particular for the electrostatic potential, than in the high shear simulation.

The poloidal and toroidal spectra of fluctuations on a flux surface close to the LCFS are shown in Fig. \ref{fig:fourier_spectrum_tokamak} for the low and high magnetic-shear simulations. The dominant modes are such that $n/m=1/2$ (in the low-shear simulation) and $n/m=3/8$ (in the high-shear simulation). These values correspond to $1/q$ around the steepest part of the equilibrium pressure, implying that $\kpar=(B^{\varphi}/B)(m/q-n)$ of the dominant mode approximately vanishes in both simulations, and thus modes are field-aligned. Moreover, the average phase shift between density and potential fluctuations is around $\pi/2$ with $\Tilde{\Phi} \gg \Tilde{N}$ and the density and $E \times B$ fluctuations show large amplitude on the low-field side. These features suggest that turbulence originates from curvature-driven instabilities, such as BMs. \textcolor{red}This is also verified by tests where we zero out the interchange
drive, i.e. the curvature term in Eq. (\ref{eq:GBSomega}), $C(pe + \tau pi)$, and we observe a significant steepening of the pressure profile. Indeed, previous studies of SOL turbulent regimes in limited tokamaks  \citep{annamaria_turbulent_regimes} predict that turbulence is driven by resistive BMs at the considered values of $\nu_0$ and $s$ for both values of shear. 

The value of $k_y\rhos$ of the dominant mode, which is proportional to $m$, is significantly larger in the high-shear case, as Fig. ~\ref{fig:fourier_spectrum_tokamak}  shows, with $m = 8$ in the high-shear simulation and $m = 2$ in the low-shear case. In the low-shear results, we observe that the poloidal extension of the structures is of the order of the radial extension and of $L_p$, which is comparable to $a$, i.e. $k_y\sim k_x \sim 1/L_p$, and the relation $k_x \ll k_y$ does not apply. 

We further note that a larger number of modes are excited in the high-shear case than in the low-shear. To explain this, we note that the growth rate of BM peaks at $\kpar \approx0$, implying $m/n \sim q$. In the low-shear case, since $q(r) \approx q_0$ with $q_0$ for all $r$, only one combination of $(m, n)$, up to a common multiplicative factor for $m$ and $n$ providing the same $m/n$, makes $\kpar$ minimal. On the other hand, in the high-shear case, a larger number of combinations of $m/n$ minimise $k_\parallel$ given that $q$ is a function of the radial position and thus crosses several rational numbers. Therefore, a large number of modes, possibly developing into broad-band turbulence, are expected to be present in a high-shear configuration. Indeed, the simulation results show that fluctuations in the low shear case are coherent, with the dominant $m = n q_0$ mode across the radial profile; while in the high-shear case the dominant $m$ and $n$ vary across the radial profile to satisfy the field-aligned property. Among the possible combinations of $m/n$, the simulation usually presents the lowest possible $m$ value, which leads to larger transport as this scales typically with $m^{-2}$ as shown in \citet{zeiler}.

\begin{figure*}
\centering
 \includegraphics[scale=0.35]{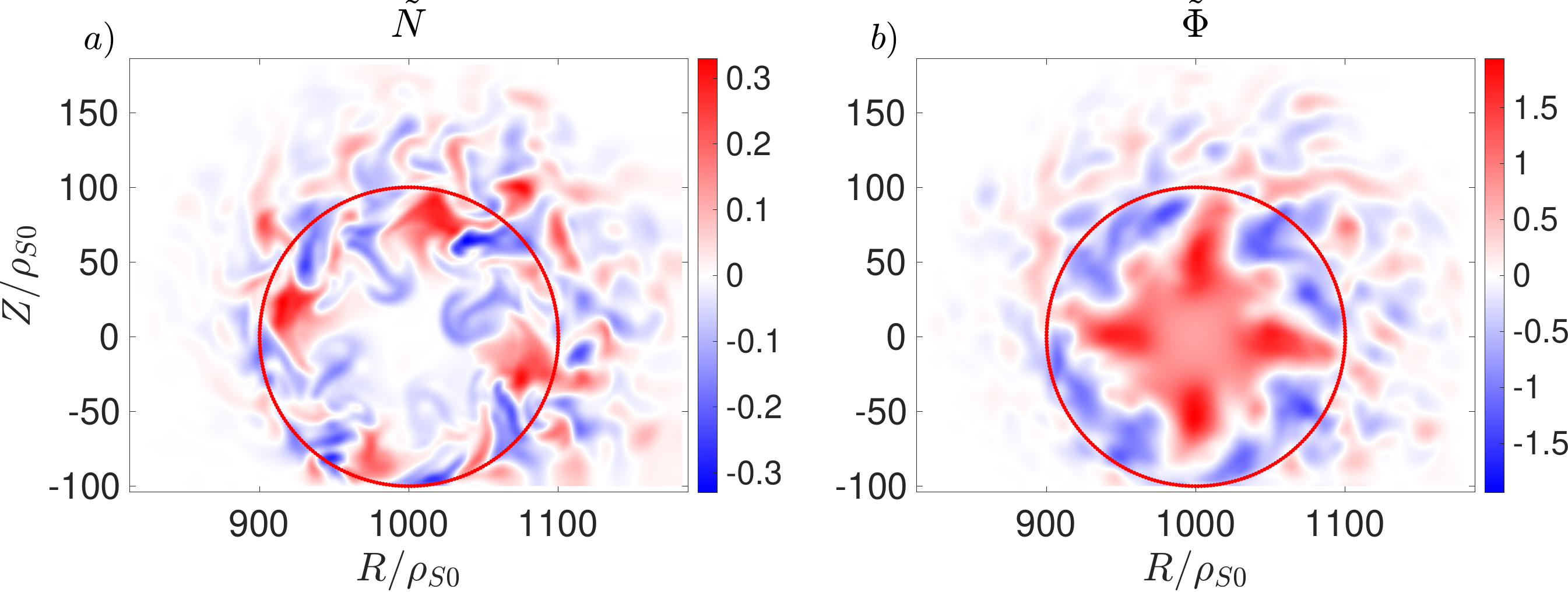}
 \caption{Density and potential fluctuations at low shear when $q_0=4$ and $q_a=4.1$.}
 \label{fig:n_strmf_fluctuations_q_4}
\end{figure*}
Similar results to the high-shear case have been obtained with $q_0=1$ and $q_a =4$, and with $q_0 = 2$ and $q_a = 4$ (data not shown). We remark that the turbulent modes are coherent in the low-shear regime for larger safety factor values. Indeed, simulations with $q_0=4$ and $q_a=4.1$ reveal that the potential presents a large and radially elongated mode with $m=4$ (see in Fig.~\ref{fig:n_strmf_fluctuations_q_4}). In addition, when a flat q-profile is considered, also with a highly irrational value of flat $q$, namely the golden ratio $ q_0 = (1+\sqrt{5})/2 \simeq 1.61$ (the hardest irrational number approximated by fractions), we observe radially and poloidally elongated structures with higher dominant mode numbers at $m/n = 13/8 \approx q_0$ (simulation results not shown). In this case, the fluctuation spectrum still shows coherent features, similar to those in Figure \ref{fig:fourier_spectrum_tokamak} $a)$, approximatively independent of the radial profile.  These results demonstrate that a low magnetic shear significantly influences the properties of tokamak edge turbulence, providing a new regime with coherent features such that $k_x \sim k_y \sim 1/L_p$, which has not been reported by previous simulations. 

We remark that in \citet{giacomin2022turbulent}, a regime where $1/k_x \sim L_p$ has been related to pressure profile collapse and the crossing of the density limit. However, our simulations consider values of $\nu_0$ and density that are far away from this limit. Moreover, $k_y \ll k_x$ was still considered in the analysis of the degraded confinement regime studied by \citet{giacomin2022turbulent}. 
Finally, we note that in the MST employed as a tokamak, characterised by a low-shear, low-$\beta$ configuration, interchange-like modes are also indicated as possible instabilities \cite{hurst2022self, hurst2024tokamak}. More recently, large "snake structures" dominating the dynamics with $m = n = 1$ have been observed in MST\cite{schmall2025characterization} for $q$ at the edge larger than $2.2$ with $q$ at the axis 1.


\section{Global linear theory for ballooning modes} \label{Sec:Linear}

We now show that the differences between low shear and high shear can be explained using a linear model. Since turbulence is driven by ballooning instabilities in the simulations presented here, we consider a reduced model from Eqs.~(\ref{eq:GBSn}-\ref{eq:GBS_phi}) that includes the main elements of the ballooning instability~\citep{annamaria_linear_theory, halpern2013ideal}. First, we consider the cold ion limit, $T_i = 0$, and we neglect sound waves, assuming $k_\parallel \ll \partial_t $ (or $c_{s0} k_\parallel\ll \partial_t$ in physical units). We also neglect drift waves, with the hypothesis $\omega_{de} \ll  \partial_t$, where $\omega_{de} = \textbf{k} \cdot \textbf{V}_{d e}$ is the diamagnetic frequency, and $\textbf{V}_{d e} = \nabla N T_e \times \textbf{b}/(e N B)$ is the diamagnetic velocity. Finally, we neglect compressibility terms associated with the curvature effects in the density and
temperature equations. As a result, the electrons and ions dynamics are decoupled, and, focusing on the electrons species (subscripts are dropped in the following for clarity), we consider:  
\begin{align}
    &\pt{N} = -\frac{\rorhos}{B}\left[\Phi,N\right], \label{eq:linN}\\ 
    &\pt{T} = -\frac{\rorhos}{B}\left[\Phi,T\right], \label{eq:linT} \\ 
    &\pt{V} = -\nu\left(\frac{m_i}{m_e}\right)N V + \left(\frac{m_i}{m_e}\right)\nablapar\Phi ,\label{eq:linV} \\ 
    & \frac{\p\nablaperp^2\Phi}{\p t} = -\frac{\rorhos}{B}\left[\Phi,\nablaperp^2\Phi \right] -\frac{B^2}{N}\nablapar(N V) + \frac{2B}{N}C(N T). \label{eq:linPhi}
\end{align}
The BM drive is given by the curvature-driven term $C(N T)$ in Eq. (\ref{eq:linPhi}). In the $(r, \varphi, \theta)$ coordinate system and expanding in the large aspect ratio, the normalised geometrical operators in Eqs. (\ref{eq:parallel_gradient}-\ref{eq:perpendicular_laplacian}) become:
\begin{align}
    &\nabla_{\parallel} f = \frac{1}{q} \frac{\partial f}{\partial \theta} + \frac{\partial f}{\partial \varphi}, \label{eq:pargrad}\\
    & [\Phi, f] = \frac{1}{r}\Bigg(\frac{\partial \Phi}{\partial r}\frac{\partial f}{\partial \theta} -  \frac{\partial \Phi}{\partial \theta}\frac{\partial f}{\partial r}\Bigg)\\
    & C(f) = \sin\theta \frac{\partial f}{\partial r} + \frac{\cos \theta}{r} \frac{\partial f}{\partial \theta} \\
    & \nabla_{\perp}^2 f = \frac{\partial^2 f}{\partial r^2} + \frac{1}{r}\frac{\partial f}{\partial r} + \frac{1}{r^2}\frac{\partial^2 f}{\partial^2 \theta}. \label{eq:perplap}
\end{align}

We linearise Eqs. (\ref{eq:linN}- \ref{eq:linPhi}) by expressing a field $f$ as $f(t, r, \theta, \varphi) =  f_0(r) + \Tilde{f}(t, r, \theta, \varphi)$ with $\tilde{f} = \sum_n e^{\sigma_n t} e^{-i n \varphi} f_n(r, \theta)$, where $f_0$ is the equilibrium background, $\tilde{f} \ll f_0$ its perturbation, $\sigma_n = \gamma_n + i \omega_n$ with $\gamma_n$ the growth rate and $\omega_n$ the real frequency, and the perturbation $\tilde{f}$ is decomposed toroidally into Fourier modes, with $n$ the toroidal mode number. For simplicity, we drop the subscript $n$ from $f$, $\gamma$ and $\omega$ in the following. We set $\Phi_0 = 0$ which is equivalent to analyse the mode evolution in the reference frame rotating poloidally with the $E\times B$ drift. We neglect parallel velocity shear effects imposing $V_0 = 0$. Moreover, we assume $n_0 = T_0 = f_0(r)$, with $f_0$ given by
\begin{align} 
      f_0(r) = \frac{1}{h} \log \Bigg[\tanh(\frac{\alpha - r}{L_0}) + f_{00}\Bigg],
\end{align}
where
\begin{align}
      & \alpha =  \hat{r} + \frac{L_0}{4} \log\frac{f_{00} - 1}{f_{00} + 1},
\end{align}
having defined
\begin{align}
    h = \log\left[f_{00} +  \tanh\left(\frac{\hat{r}}{L_0} + \frac{1}{4} \log \frac{f_{00} - 1}{f_{00} + 1}\right)\right],
\end{align}
We remark that $f_0^{'}$ peaks at $r = \hat{r}$, the parameter $L_0$ controls the characteristic gradient length scale of $f_0$, and the constant $f_{00}$ is a dimensionless parameter that sets the background value. In the present analysis we consider different values of $L_0$, from $L_0/a = 0.03$ corresponding to a localised, large equilibrium gradient, to $L_0/a = 0.3$ corresponding to a distributed, small equilibrium gradient. We take $\hat{r} = a/2$, to center the profile in the radial direction, and $f_{00} = 2$, and we refer to the values of magnetic shear and safety factor at $\hat{r}$ as $\hat{s}$ and $\hat{q}$, respectively. We center the equilibrium profile in the domain to improve smoothness at $r = 0$ and $r = a$.

With these hypotheses, we linearise Eqs. (\ref{eq:linN}-\ref{eq:linPhi}) with the geometrical operators given in Eqs. (\ref{eq:pargrad}-\ref{eq:perplap}), yielding 
\begin{align}
        & \sigma N = \sigma T = \Bigg[- \rho_{*}^{-1}\frac{ f_0^{'}}{r}  \frac{\partial}{\partial \theta}\Bigg] \Phi,\label{Eq:Eign}\\
        & \sigma V = \Bigg[- \frac{m_i}{m_e}\nu_0 \frac{1}{f_0^{1/2}}\Bigg] V + \frac{m_i}{m_e}\Bigg[\frac{1}{q}\frac{\partial}{\partial \theta} - i n\Bigg]\Phi, \label{Eq:EigV}\\ \notag 
        & \sigma \Bigg[\frac{\partial^2}{\partial r^2} + \frac{1}{r} \frac{\partial}{\partial r} + \frac{1}{r^2} \frac{\partial^2}{\partial \theta^2}\Bigg] \Phi = \Bigg[ - 2 \frac{\sin \theta}{f_0}f_0^{'} + 2 \sin \theta \frac{\partial}{\partial r}  \\ 
        & + 2 \frac{\cos \theta}{r} \frac{\partial}{\partial \theta} \Bigg] N  + \Bigg[2 \frac{\sin \theta}{f_0}f_0^{'} + 2 \sin \theta \frac{\partial}{\partial r} +  2 \frac{\cos \theta}{r} \frac{\partial}{\partial \theta}\Bigg] T \notag  \\ 
        & - \Bigg[\frac{1}{q}\frac{\partial}{\partial \theta} - i n\Bigg] V\label{Eq:EigPhi},
\end{align}
where the prime stands for the radial derivative.

In the following, we exploit the field-aligned nature of the ballooning mode \citep{annamaria_linear_theory} to simplify the eigenvalue problem in Eqs. (\ref{Eq:Eign}-\ref{Eq:EigPhi}). Since $k_\parallel \sim  m/q - n$, the radial profile of $q$ has an impact on the field-aligned structure of the mode. Hence, we consider separately a constant and a varying $q$ along $r$, i.e. vanishing and non-vanishing profiles of magnetic shear, respectively.

We first consider the non-vanishing shear case. We use the ballooning transform \citep{connor1978shear, connor2004stability}, expressing a toroidal mode $\tilde{f}(r, \theta)$ as:
\begin{align}\label{eq:ballonigrep}
    \tilde{f}(r, \theta) = \sum_{m = - \infty}^{+ \infty} e^{- i m \theta} \int_{- \infty}^{+\infty} e^{i m \eta} \hat{f}(r, \eta) d\eta,
\end{align}
by performing a transformation from the periodic coordinate $\theta \in [0, 2\pi)$ to $\eta \in (-\infty, +\infty)$. As showed in \cite{connor1979high}, the transformation in Eq. (\ref{eq:ballonigrep}) preserves the spectrum of the eigenvalue problem in Eqs. (\ref{Eq:Eign}-\ref{Eq:EigPhi}) assuming that $\hat{f}$ vanishes for $\eta \to \pm \infty$. By construction, $\hat{f}$ is not periodic and, therefore, each field can be expressed in an eikonal form 
\begin{align}\label{eq:Eikonalform}
    \hat{f}(r, \eta) = u_f(r, \eta)e^{i S},
\end{align}
where $S$ is the fast-varying eikonal function, and $u_f$ is the slowly oscillating envelope, so that $\nabla_\parallel S = 0$. In our case, considering the toroidal phase component of the mode, the eikonal function is $S = n[\varphi + q(r) \eta + \eta_0]$. The value $\eta_0$ represents the poloidal angle where the perturbation has a local maximum amplitude, being a free parameter of the problem. Given the poloidal symmetry of Eqs. (\ref{Eq:Eign}-\ref{Eq:EigPhi}), this is either $\eta_0 = 0$ or $\eta_0 = \pi$. A numerical investigation on the influence of $\eta_0$ on $\gamma$ shows that the growth rate is maximum for $\eta_0 = 0$, consistently with the ballooning instability peaking at the low field side. Hence, we take $\eta_0 = 0$ in the following.

\begin{figure*}
 \centering
 \includegraphics[scale=0.35]{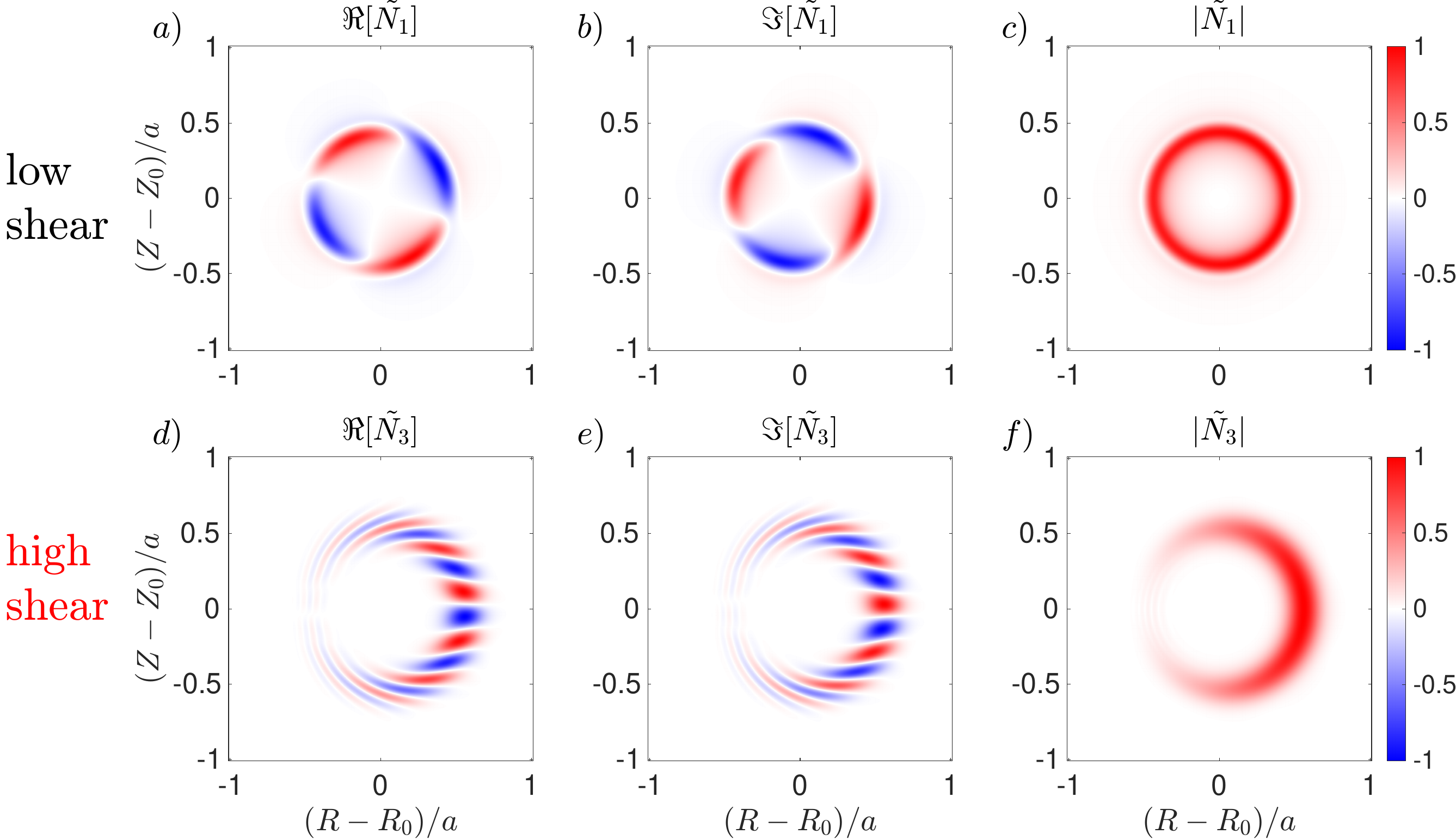}
 \caption{Radial-poloidal mode structure for fastest growing BM in the low shear case ($\hat{s} = 0, \hat{q} = 2.0$) with $n = 1$ (top) and in the high shear ($\hat{s} = 1.6, \hat{q} = 2.0$) with $n = 3$ (bottom). Real (left), imaginary (middle) parts and module (right) are considered.}
 \label{fig:struct2D}
\end{figure*}

The reconstruction of the mode is then performed according to \citet{taylor2015special}, that is
\begin{align} 
    \tilde{f}(r, \theta) = \sum_{M} e^{-i n q(\theta -  2\pi M)} u_f(r, \theta - 2 \pi M),
\end{align}
 
Expressing $N$, $T$, $V$ and $\Phi$ with the ballooning transformation in Eq. (\ref{eq:ballonigrep}) in the eigenvalue problem in Eqs. (\ref{Eq:Eign}-\ref{Eq:EigPhi}), and employing the eikonal form in Eq. (\ref{eq:Eikonalform}), we obtain:
\begin{align}
    &\sigma N = \sigma T = \Bigg[- \rho_{*}^{-1}\frac{ f_0^{'}}{r}\Bigg( \frac{\partial}{\partial \eta} - i n q \Bigg)\Bigg] \Phi \label{eq:highN&T}\\
    &\sigma V =  \Bigg[- \frac{m_i}{m_e}\nu_0 \frac{1}{f_0^{1/2}}\Bigg] V + \frac{m_i}{m_e}\Bigg[\frac{1}{q}\frac{\partial}{\partial \eta}\Bigg]\Phi \label{eq:highV}\\
    & \sigma \Bigg[\frac{\partial^2}{\partial r^2} + \frac{1}{r} \frac{\partial}{\partial r} + \frac{1}{r^2} \frac{\partial^2}{\partial \eta^2}  - 2i n \eta q^{'} \frac{\partial}{\partial r} - 2 i n \frac{q}{r^2} \frac{\partial}{\partial \eta}  -  i n \frac{q^{'}}{r} \eta  \notag \\
    & - n^2 \eta^2 (q^{'})^2  - in \eta q^{''} - n^2 \frac{q^2}{r^2}\Bigg] \Phi \notag \\
    & = 2\Bigg[ -  \frac{\sin \eta}{f_0}f_0^{'} +  \sin \eta \frac{\partial}{\partial r} + \frac{\cos \eta}{r} \frac{\partial }{\partial \eta}
    - i n \sin \eta q^{'} \eta - i n \frac{\cos \eta }{r} q \Bigg] N \notag \\
    &+2\Bigg[ \frac{\sin \eta}{f_0}f_0^{'} + \sin \eta \frac{\partial}{\partial r} + \frac{\cos \eta}{r} \frac{\partial }{\partial \eta}  - i n \sin \eta q^{'} \eta - i n \frac{\cos \eta }{r} q \Bigg] T  \notag \\
    & - \Bigg[\frac{1}{q}\frac{\partial }{\partial \eta}\Bigg] V. \label{eq:highPhi}
\end{align}
To simplify the notation, in Eqs. (\ref{eq:highN&T}-\ref{eq:highPhi}) we denote the envelope of the eikonal part of the ballooning representation, $u_f$ in Eq. (\ref{eq:Eikonalform}), simply with the corresponding field $f$, i.e. indicate $u_f$ with $f$. We remark that, in this work, the ballooning representation is not used for decoupling parallel and perpendicular dynamics by taking the ideal ballooning limit of $n \to \infty$, as it is classically done in magnetohydrodynamics (MHD) \citep{connor1978shear}. Here, we only use the ballooning transform as a representation, similarly to the approach used in PEST-II \citep{dewar1981n}.

In the limit of low shear and low $n$, the ballooning representation breaks down due to the fact that the solutions $u$, and $\hat{f}$, are not localised in $\eta$. This problem is well known in MHD where the criterion of the validity of such representation is $n \gg (\psi q^{'})^{-2}$ , where $\psi$ is the poloidal magnetic flux, making the vanishing shear case pathological\citep{hastie1981validity}. Even though the identification of a similar criterion for Eqs. (\ref{eq:linN}-\ref{eq:linPhi}) goes beyond the scope of this work, an initial numerical investigation suggests that the representation in Eq. (\ref{eq:ballonigrep}) is not suited for investigations in the low-shear regime considered in this paper.

For the low-shear cases we proceed as explained in \citet{hastie1981validity}. We consider the following representation for the fields in Eqs. (\ref{Eq:Eign}-\ref{Eq:EigPhi}):
\begin{align}\label{Eq:lowshearrep}
    \tilde{f}(r, \theta, \phi) = \hat{f}(r, \theta) e^{i(m\theta - n \varphi)},
\end{align}
where $m = n q_0$. Using the representation in Eq. ($\ref{Eq:lowshearrep}$) into the eigenvalue problem in Eqs. (\ref{Eq:Eign}-\ref{Eq:EigPhi}), we obtain:
\begin{align}
        & \sigma N = \sigma T = \Bigg[- \rho_{*}^{-1}\frac{ f_0^{'}}{r} \Bigg(\frac{\partial}{\partial \theta} - i m \Bigg) \Bigg] \Phi ,\label{Eq:EigZeron}\\
        & \sigma V = \Bigg[- \frac{m_i}{m_e}\nu_0 \frac{1}{f_0^{1/2}}\Bigg] V + \frac{m_i}{m_e}\Bigg[\frac{1}{q}\frac{\partial }{\partial \theta}\Bigg]\Phi,\\ \notag 
        & \sigma \Bigg[\frac{\partial^2}{\partial r^2} + \frac{1}{r} \frac{\partial}{\partial r} - \frac{m^2}{r^2} \Bigg] \Phi = \Bigg[ - 2 \frac{\sin \theta}{f_0}f_0^{'} + 2  \sin \theta \frac{\partial}{\partial r}   \notag \\
    &+ 2 \frac{\cos \theta}{r} \Bigg(\frac{\partial}{\partial \theta} - i m\Bigg) \Bigg] N \notag  \Bigg[2 \frac{\sin \theta}{f_0}f_0^{'} + 2 \sin \theta \frac{\partial}{\partial r}  \notag \\
    & +  \frac{\cos \theta}{r} \Bigg(\frac{\partial}{\partial \theta} - i m\Bigg) \Bigg] T 
         - i\Bigg[\frac{1}{q}\frac{\partial }{\partial \theta}\Bigg] V\label{Eq:EigZeroPhi}. 
\end{align}
\begin{figure*}
 \centering
 \includegraphics[scale=0.5]{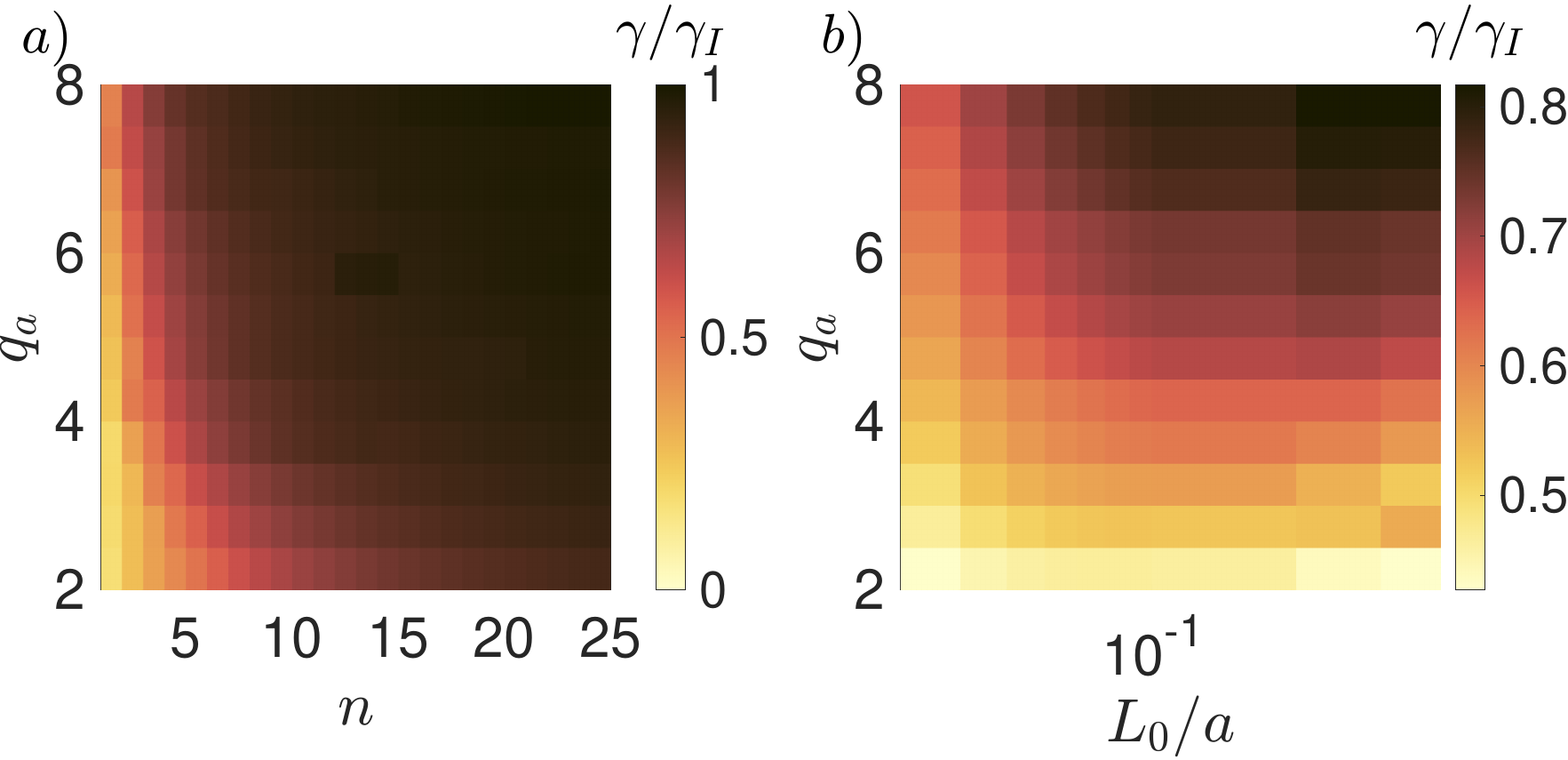}
\caption{Largest growth rate as a function of safety factor at the edge $q_a$ and of the mode number $n$ in $(a)$ and as a function of $q_a$ and $L_0$ in $(b)$. In $a)$, $L_0 = 0.2a$, while} in $b)$ $n$ is fixed to 4. In both cases, $q_0 = 2$.
\label{fig:growthrate}
\end{figure*}

To simplify the notation also in this case, we drop the hat, which is present in Eq. (\ref{Eq:lowshearrep}), in Eqs (\ref{Eq:EigZeron}-\ref{Eq:EigZeroPhi}). The two representations in  Eqs.(\ref{Eq:lowshearrep}) and (\ref{eq:ballonigrep}) allow solving the dynamics without fast oscillating modes in the poloidal plane, using lower resolution and, thus, accelerating numerical convergence. 

A numerical implementation of the three eigenvalue problems, Eqs. (\ref{Eq:Eign}-\ref{Eq:EigPhi}), (\ref{eq:highN&T}-\ref{eq:highPhi}) and Eqs. (\ref{Eq:EigZeron}-\ref{Eq:EigZeroPhi}), is based on a finite difference scheme with an iterative solver that makes use of a Krylov-Schur algorithm for large sparse eigenvalue problems \citep{stewart2002krylov} implemented by \texttt{eigs} function of \texttt{MATLAB} \citep{MATLAB:2024a}. Details are reported in Appendix \ref{Sec:numimplementationEig}. In our analysis, we consider the eigenvalue (and their corresponding eigenvectors) with the largest growth rate. 

The eigenvalue and eigenmodes computed via the low-shear representation in Eq. (\ref{Eq:lowshearrep}) and via the ballooning representation in Eq. (\ref{eq:ballonigrep}) for high-shear cases are successfully verified by benchmarking these with the direct solution of Eqs. (\ref{Eq:Eign}-\ref{Eq:EigPhi}) in fully converged cases for $n\lesssim 12$, showing that these representations correctly capture the BM dynamics in the corresponding shear regimes. Therefore, if not specified otherwise, in the following we compute vanishing shear results solving Eqs. (\ref{Eq:EigZeron}-\ref{Eq:EigZeroPhi}), and for non-vanishing shear cases, we solve Eqs. (\ref{eq:highN&T}-\ref{eq:highPhi}). The results in the following sections are obtained by fixing the value of $q_0 = 2$ and increasing $q_a$, effectively increasing the local values of $q$ and $s$ at the peak of the equilibrium gradient. If not specified otherwise, we fix $m_i/m_e = 200$, $\nu_0 = 0.1$, $\rho_*^{-1} = 1000$, $a = 100\, \rho_{s0}$, and $L_0 = 0.2 a$.

\begin{figure}[h]
 \centering
 \includegraphics[scale=0.3]{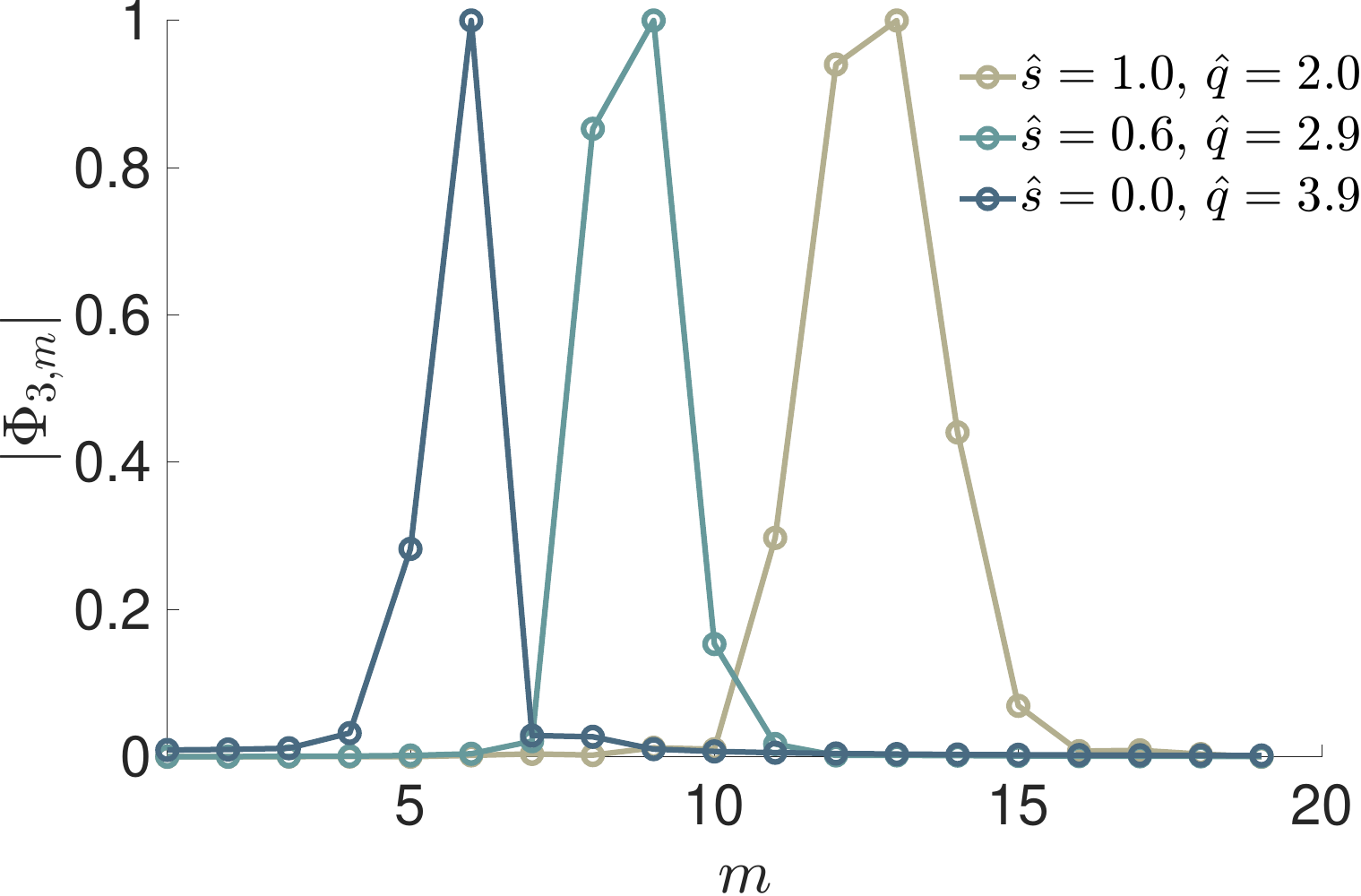}
 \caption{Poloidal spectrum of the eigenmode at $\hat{r}$ for $n = 3$ and different $\hat{s}$. Each curve is normalised so that its maximum is at 1.}
 \label{fig:polstruct}
\end{figure}
\section{Global linear structure of the ballooning mode} \label{Sec:numerciallinearres}
Figure \ref{fig:struct2D} shows the resulting eigenmodes poloidal structure for the modes dominating the non-linear simulation, $n = 3$ and $n = 1$ (as seen from Fig. \ref{fig:fourier_spectrum_tokamak}), for the high- and low-shear cases, respectively. In the high shear case, the values of $q_0$ and $q_a$ are such that $\hat{q}$ and $\hat{s}$ at $a/2$ are close to the corresponding ones at the edge in the nonlinear simulations. The low-shear eigenmode is highly coherent, with a dominant $m = 2$ mode with no clear localisation around $\theta = 0$. On the other hand, the high-shear eigenmode reveals a filamentary structure localised on the low-field side with high poloidal modes coexisting with a dominant $m = 8$ mode. In both cases, the dominant poloidal mode number is consistent with the non-linear simulation results. In addition, the mode is radially localised around the peak of the equilibrium gradient, and the eigenvalue of the fastest growing mode is real, consistently with previous studies on resistive ballooning instability in the tokamak edge that points to a purely growing mode \citep{annamaria_turbulent_regimes, halpern2013ideal} for both the high- and low- shear cases. The mode structure changes, from $k_y \lesssim k_x$ in the low-shear to $k_x \ll k_y$ in the high-shear case. We note that, the influence of the shear on the radial elongation and poloidal mode structure has also been observed in linear ideal and resistive infernal modes characterising hybrid tokamak scenarios \citep{graves2023non, coste2024fundamental}. Large-scale modes in low-shear configurations are also observed in linear MHD studies of the MST tokamak at $\beta \sim 5\times 10^{-3}$, pointing to resistive wall tearing modes \citep{strauss2023mst}.

\begin{figure}[h]
 \centering
 \includegraphics[scale=0.3]{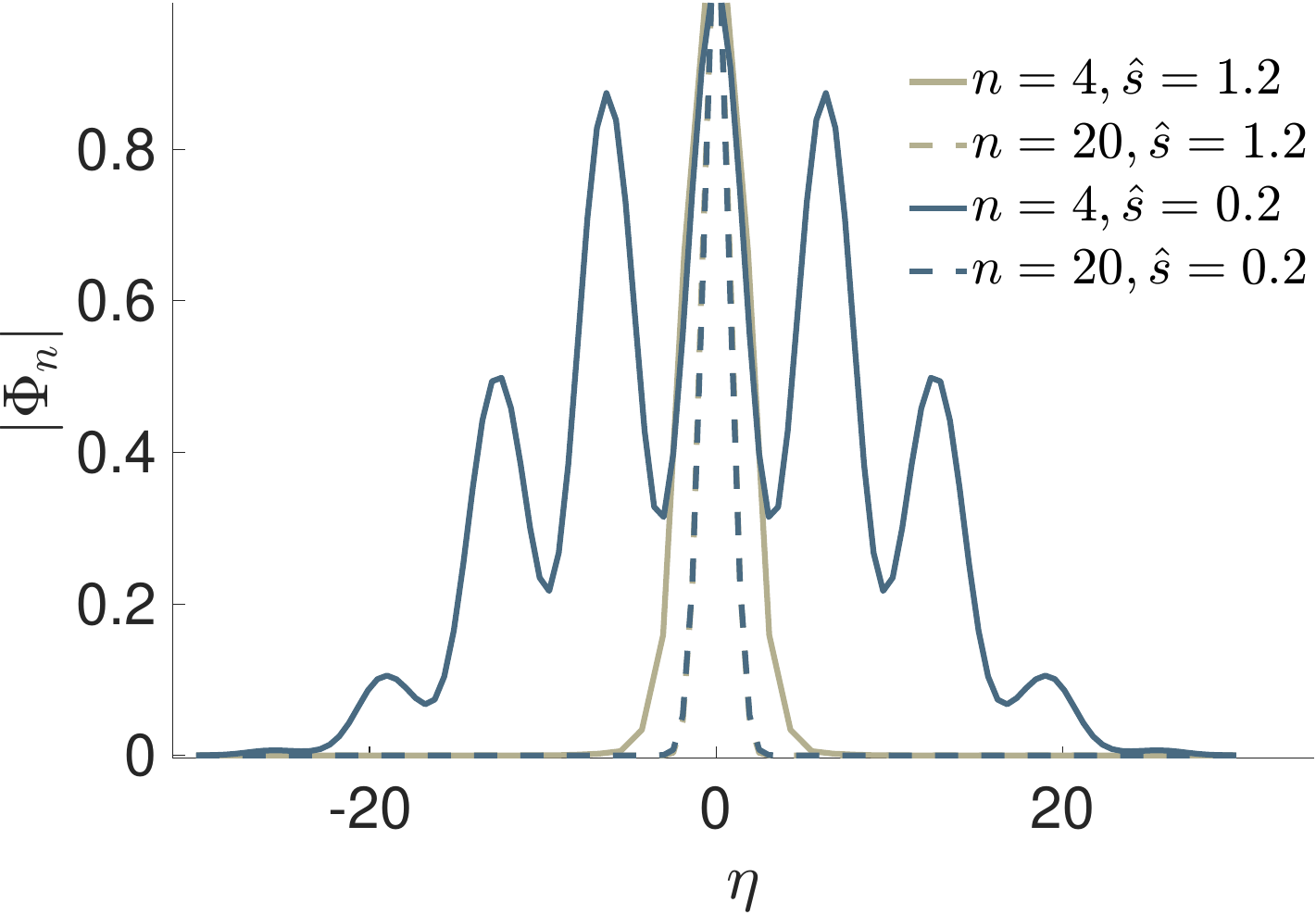}
 \caption{Structure of the eigenmode along the ballooning angle $\eta$ for different $n$ and $\hat{s}$. Each curve is normalised so that its maximum is at 1. }
 \label{fig:ballooningstruct}
\end{figure}
The numerical solution shows that the most unstable mode is destabilised by increasing values of $q_a$, $n$ and by the steepening of the equilibrium profile measured by $L_0/a$. This effect is discussed in Figure \ref{fig:growthrate}, where the growth rate $\gamma$ is shown as a function of $n$ (Fig. \ref{fig:growthrate}$ a)$ and of $L_0/a$ in (Fig. \ref{fig:growthrate} $b)$. The growth rate is normalised to the interchange growth rate, $\gamma_I = \sqrt{2 R_0/L_0}$ \citep{annamaria_linear_theory}, valid in the $k_y\to \infty$ limit. We observe that, indeed, $\gamma \to \gamma_I$ in the limit of high $n$ and $q_a$. We also observe that $\gamma$ increases with larger resistivity and $1/L_0$, consistently with previous studies\cite{annamaria_linear_theory}.

We investigate the poloidal mode structure by Fourier transforming the poloidal profile of $|\Phi|$ and  obtaining the set poloidal mode amplitudes $|\Phi_m|(r)$. We observe that the $q$-profile determines the dominant poloidal mode number according to the relation $m \simeq n q$, underlying the field-aligned character of the ballooning instability. In the low-shear case, $m$ is constant along the radial direction, providing a coherent mode. On the other hand, in the high-shear case, $m$ varies radially, resulting into a filamentary shape with a broader poloidal spectra. In this case, the poloidal mode amplitude peaks at the rational surface closest to $\hat{r}$. This difference between the poloidal spectrum of the high- and low-shear cases is depicted in Fig. \ref{fig:polstruct}, where $|\Phi_m(\hat{r})|$, is plotted for different shear values.

\begin{figure*}
 \centering
 \includegraphics[scale=0.35]{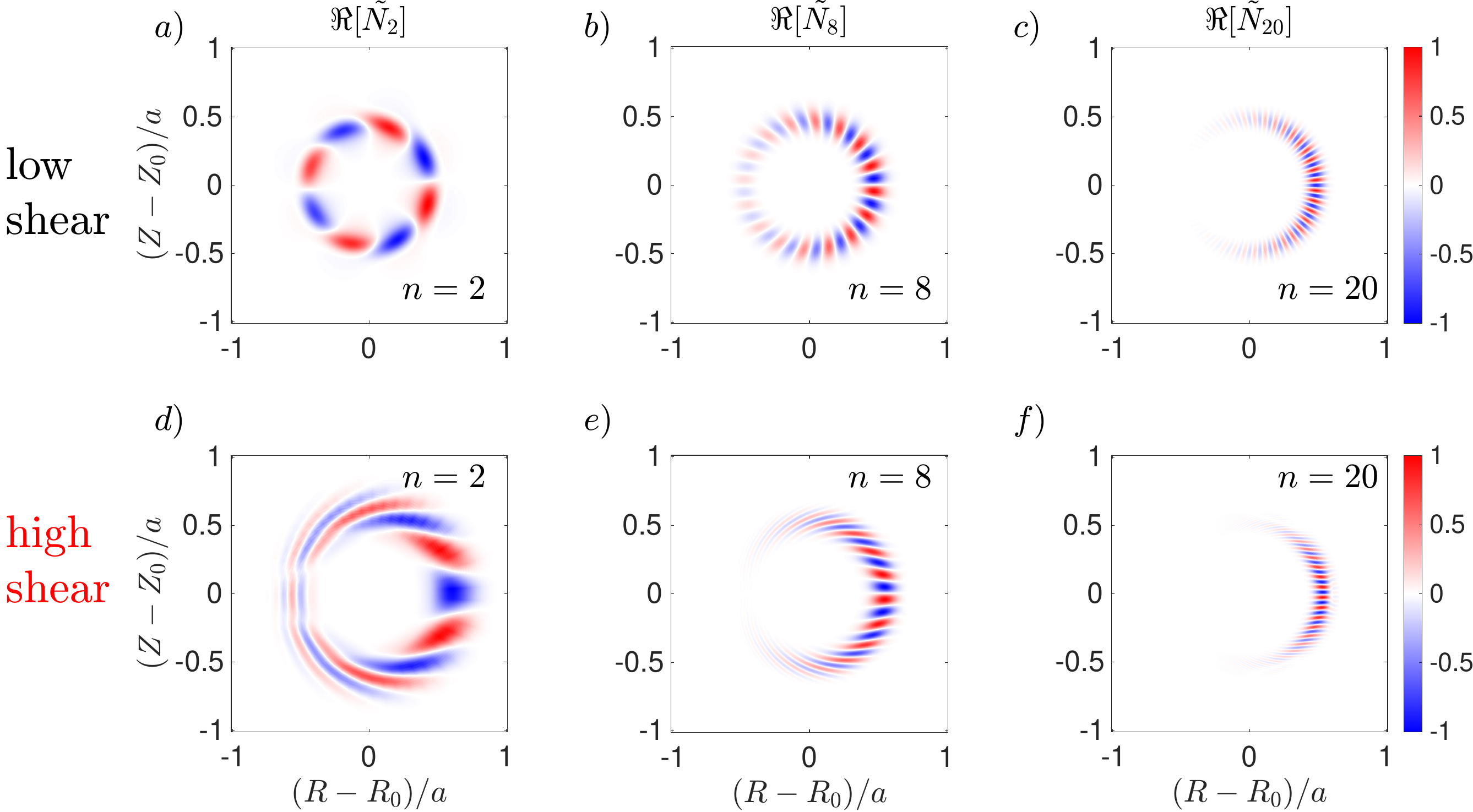}
 \caption{Real part of the eigenmode for low shear in $(a)$, $(b)$ and $(c)$ with $\hat{s} = 0, \hat{q} = 2.0$, and high shear in $(d)$, $(e)$, and $(f)$ with $\hat{s} = 1.6, \hat{q} = 2.0$. The toroidal mode numbers considered are $n = 2, 8, 20$, respectively from left to right.}
 \label{fig:scanstructureinn}
\end{figure*}

\begin{figure*}
 \centering
 \includegraphics[scale=0.35]{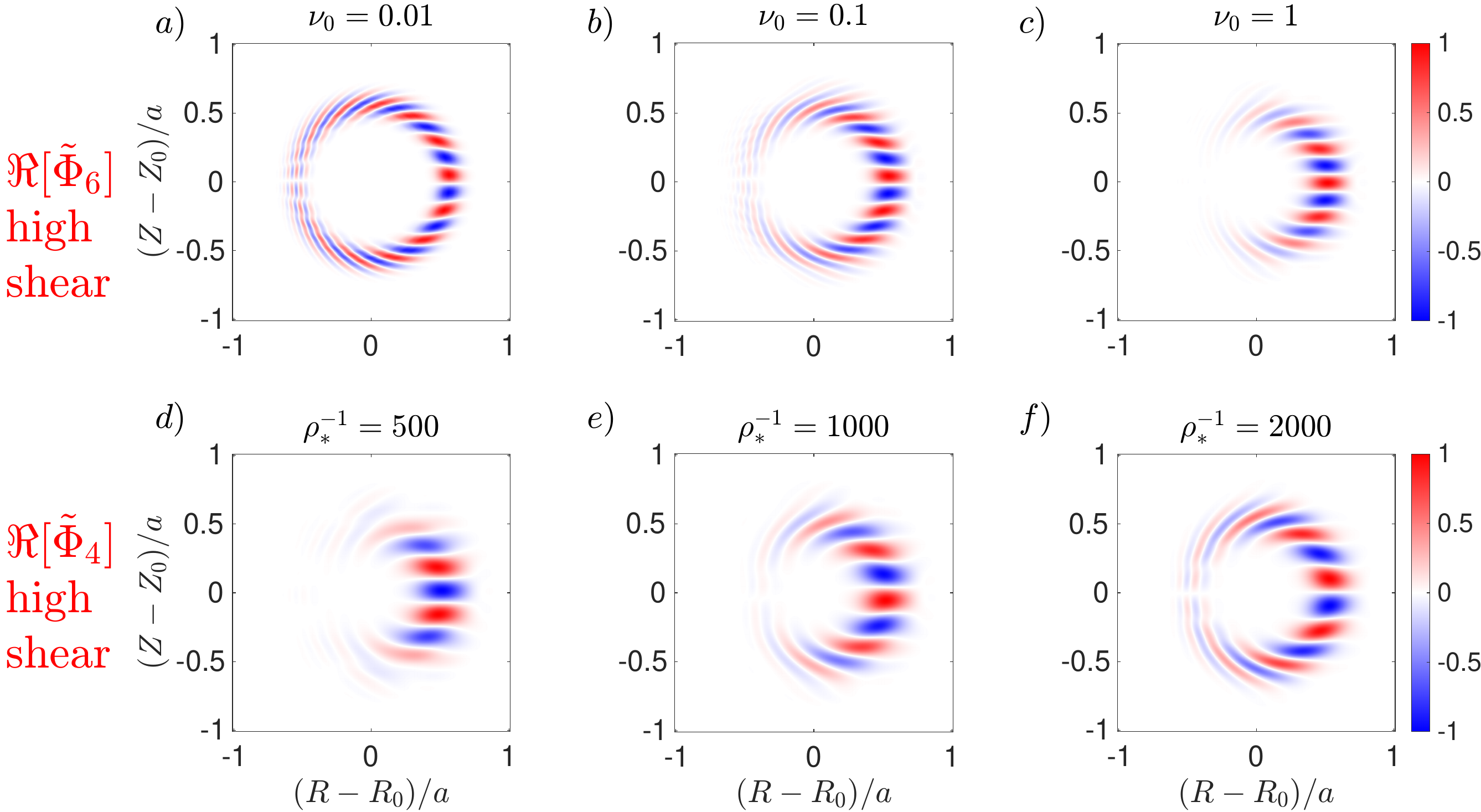}
 \caption{On the top, $\Re[\Tilde{\Phi}_6]$ for increasing values of  $\nu_0$ in a high-shear configuration with $\rho_{*}^{-1} = 1000$.
 On the bottom, $\Re[\Tilde{\Phi}_4]$ for increasing values of $\rho_*^{-1}$, while keeping $\rho_*^{-1}/a = 0.1$ in the high shear case with $\nu = 0.1$. In both scans, $q_0 = 1.0$, and $q_a = 4.0$.}
 \label{fig:structure_nu_and_size}
\end{figure*}

\begin{figure*}
\centering
 \includegraphics[scale=0.35]{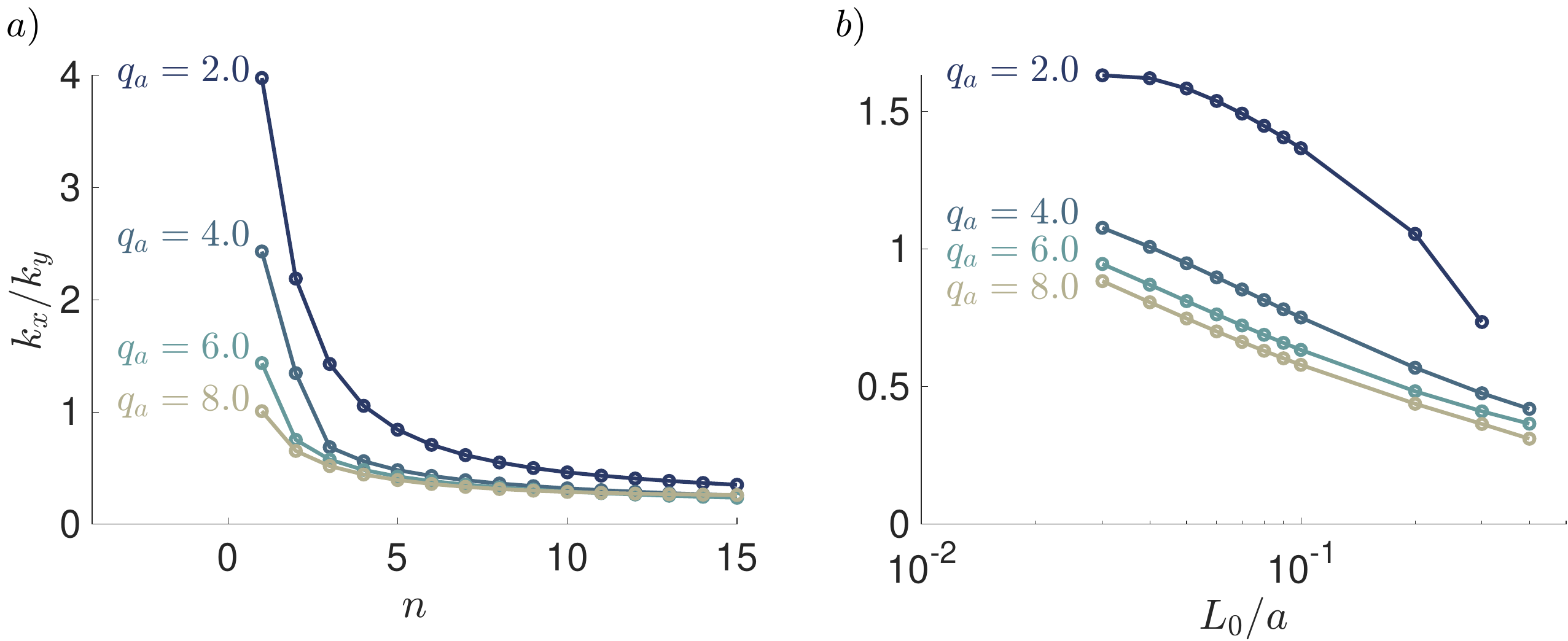}
 \caption{In $(a)$, $k_x/k_y$ as a function of $n$ for increasing $q_a$ from $2$ to $8$ with $L_0 = 0.2a$. In $(b)$, $k_x/k_y$ as a function of $L_0/a$ with $q_a$ from $2$ to $8$ with $n = 4$. In both cases $q_0 = 2$.}
 \label{fig:kxkyplot}
\end{figure*}

Increasing values of magnetic shear $\hat{s}$ and toroidal mode number enhance the poloidal mode localisation at the low-field side, analogous to the behavior of ideal BMs in MHD \citep{connor1978shear}. This effect is illustrated in Fig. \ref{fig:ballooningstruct} showing the magnitude of the eigenmodes plotted as a function of the ballooning angle $\eta$ at the radial location $\hat{r}$, corresponding to the peak of the equilibrium gradient $\hat{r}$. In addition, modes with larger toroidal numbers $n$ are more poloidally and radially localised, as it is shown in Fig. \ref{fig:scanstructureinn}, where increasing $n$ from 2 to 20 results in greater poloidal and radial localization of the eigenmode for two different values of $\hat{s}$. The ballooning character of the eigenmode is enhanced by larger values of $\nu_0$, leading to a reduction in $k_x$, causing the mode structure to extend radially, as seen on the top of Fig. \ref{fig:structure_nu_and_size}. On the other hand, more interchange-like modes are recovered by increasing $\rho_{*}^{-1}$ as showed at the bottom of Fig. \ref{fig:structure_nu_and_size}, where we perform a scan in $\rho_{*}^{-1}$ keeping fixed $a/\rho_{*}^{-1} = 0.1$. This feature highlights the curvature-driven character of the mode, since curvature scales with $1/R_0$, which corresponds to $\rho_*^{-1}$ in normalised units.

Consistently with previous studies \citep{annamaria_linear_theory}, we compute the mode poloidal wavenumber as $k_y = m q(\hat{r})/\hat{r}$. On the other hand, $k_x$ is evaluated by fitting the radial profile $|\Phi_n|$ with a Gaussian, $e^{- (r - \mu)^2/2 \chi ^2}$, at the low-field side, and finally defining $k_x = 1/\chi$ with $\mu$ being the Gaussian peak position.
Fig. \ref{fig:kxkyplot} indeed shows the resulting ratio, $k_x/k_y$,  as a function of $n$ and $q_a$ having fixed $q_0 = 2$ (Fig. \ref{fig:kxkyplot} $a$), and as a function of $L_0/a$ and $q_a$ having fixed $q_0 = 2$ (Fig. \ref{fig:kxkyplot} $b$).
Fig. \ref{fig:kxkyplot} shows that the estimate $k_x \ll k_y$ \citep{paolo_rogers_brunner_2008} is found valid for sufficiently high values of magnetic shear (high values of $q_a$) and toroidal mode number, as well as for large values of $L_0/a$. On the other hand, in the low-shear case, $k_x > k_y$, up to $n \lesssim 5$.

\section{Analytical study of the ballooning mode structure}

In this section, we provide an analytical description for the linear BM underlying the key mechanism responsible for the dependence of the mode structure on magnetic shear. To provide a unified description that is valid for all values of shear, we consider the system in Eqs. (\ref{Eq:Eign}-\ref{Eq:EigPhi}), which does not impose any functional form on $\Tilde{f}$. Reducing the problem to one equation for the electrostatic potential, decomposing in poloidal modes $\Phi = \sum_{m} \phi_m(r) e^{i m \theta}$, and projecting on a specific mode $m$, we obtain
\begin{align}\label{eq:modedescription}
    &\sigma^2 \Big(\phi_{m}^{''} + \frac{\phi_{m}^{'}}{r}\Big) + F_{m-1} \phi_{m-1}^{'} - F_{m + 1}\phi_{m+1}^{'} + G_{m}^{-}\phi_{m-1} \notag \\
    & \hspace{1cm} - G_{m}^{+}\phi_{m+1} - K_m\phi_{m} = 0,
\end{align}
with
\begin{align}\label{eq:definitioncouplingparameters}
    & F_m =  2\rho_{*}^{-1}\frac{m}{r} f_0^{'}, \quad K_m = \frac{m_i}{m_e}\frac{\sigma}{\sigma + \frac{m_i}{m_e} \frac{\nu_0}{\sqrt{f_0}}}\Bigg(\frac{m}{q} - n\Bigg)^2+ \frac{\sigma^2 m^2}{r^2} \\
    & G_m^{-}=2\rho_{*}^{-1}\frac{m-1}{r}\Bigg[ f_0^{''}  - \frac{m}{r} f_0^{'}\Bigg], \quad G_m^{+} = 2\rho_{*}^{-1}\frac{l+1}{r}\Bigg[ f_0^{''}  + \frac{m}{r}f_0^{'}\Bigg],
\end{align}
for any integer $m$. Eq. (\ref{eq:modedescription}) describes the coupling of a poloidal mode $\phi_m$ with neighbouring modes, $\phi_{m \pm 1}$. This feature, defined as poloidal mode coupling, is typical of curvature-driven instabilities in tokamaks and stellarators \citep{freidberg2014ideal}. 

\begin{figure*}
 \centering
 \includegraphics[scale=0.35]{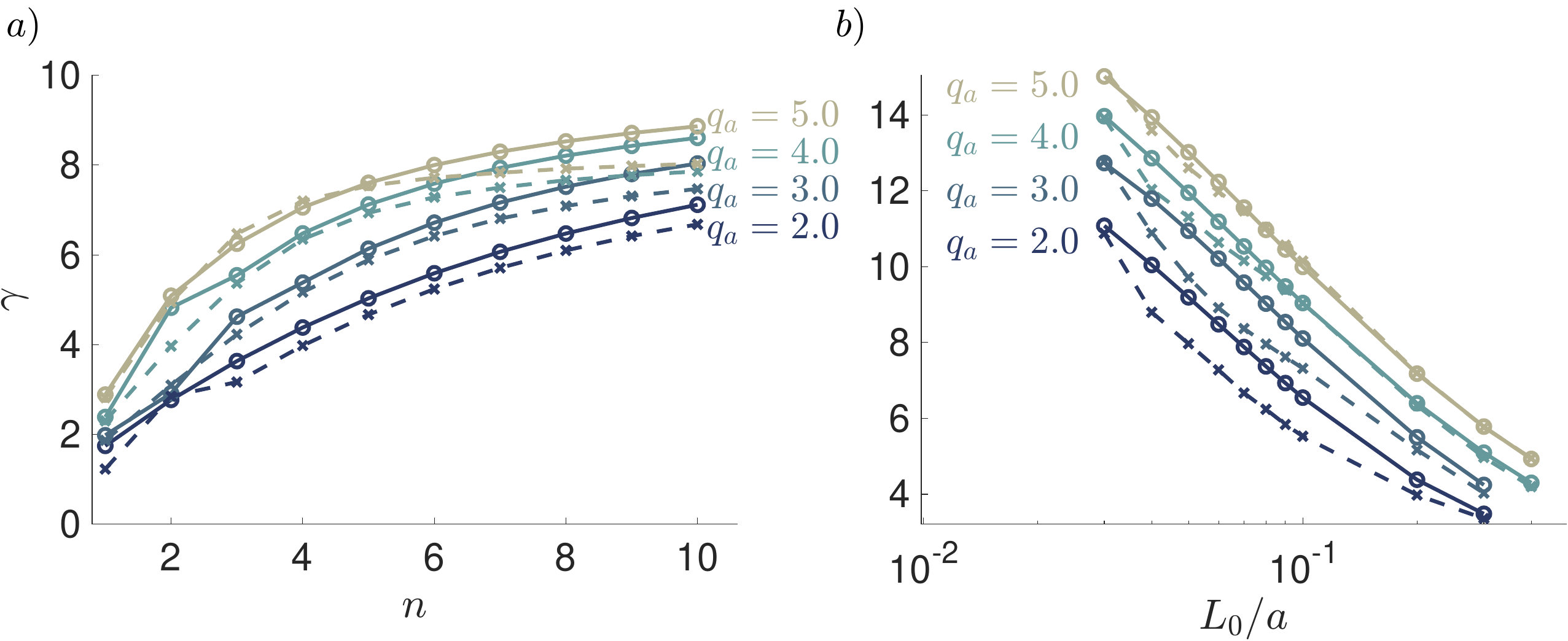}
 \caption{In $(a)$, $\gamma$ in function of $n$ with $L_0 = 0.2a$. In $(b)$, $\gamma$ as a function of $L_0/a$ fixing n = 4. In both plots, the value of $q_a$ is varied, while $q_0 = 2$ is fixed. The full lines are computed with the eigensolver, and the dashed lines only consider the coupling of the closest side bands. }
 \label{fig:solvervsanalytics}
\end{figure*}

To progress analytically, we simplify Eq. (\ref{eq:modedescription}) by
considering that only the dominant mode $m_0 = n q(\hat{r}_q)/\hat{r_q}$, with $r_q$ being the closest rational surface to $\hat{r}$, and its sidebands $m_0 \pm 1$, have finite amplitude. From Eq. (\ref{eq:modedescription}), we have the closure relation for $\phi_{m_0 \pm 1}$:
\begin{align}
    & \phi_{m_0 + 1} =  \frac{ F_{m_0} \phi_{m_0}^{'} + G^{-}_{m_0 + 1} \phi_{m_0}}{K_{m_0 + 1}}, \label{Eq:Closure+}
\end{align}
\begin{align}
     \phi_{m_0 - 1} = - \frac{F_{m_0} \phi_{m_0}^{'} + G^{+}_{m_0 - 1}\phi_{m_0}}{K_{m_0 - 1}}.\label{Eq:Closure-}
\end{align}
We remark that the approach of considering the contribution from the side-bands has been used in other contexts, such as for ideal infernal modes in MHD \citep{charlton1989resistive, coste2024fundamental}. 

Eqs. (\ref{Eq:Closure+}) and (\ref{Eq:Closure-}) show that mode coupling is induced by curvature effects via $G^{\pm}_m$ and $F_m$, and damped by finite $k_\parallel$ contributions via $K_m$. In the low-shear case, $\phi_{m_0\pm 1}$ are strongly damped by $k_\parallel$ effects since the term $(m\pm1)/q - n$ in $K_{m\pm1}$ does not vanish anywhere in $r$, resulting in a coherent mode similar to the low-shear case in Fig. \ref{fig:struct2D}. However, curvature-induced mode coupling prevails over $K_{m_0 \pm 1}$ in the high-shear case because $m\pm1/n - q$ can vanish at some radial location close to $\hat{r}$, producing an enveloped poloidal mode structure, similar to the high-shear case in Fig. \ref{fig:struct2D}.  
\begin{figure*}
 \centering
 \includegraphics[scale=0.35]{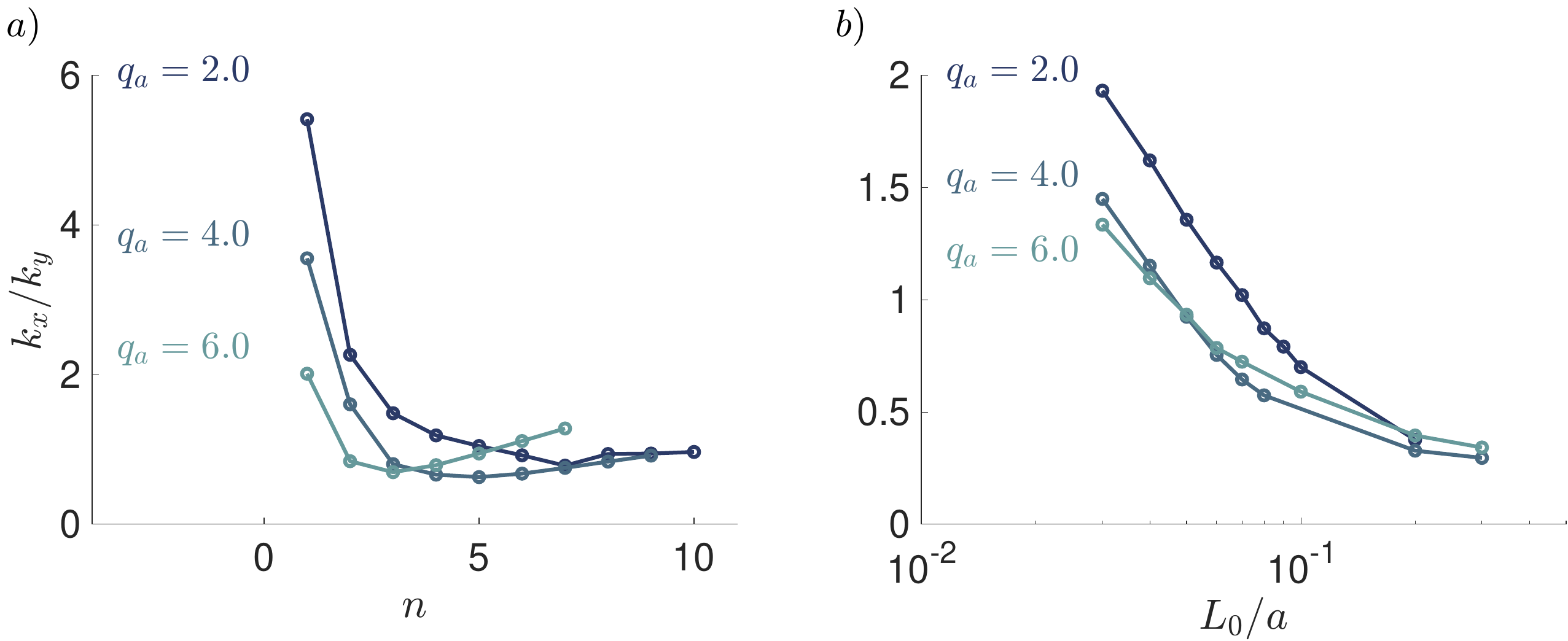}
  \caption{In $(a)$, $k_x/k_y$ in function of $n$. In $(b)$, $k_x/k_y$ as a function of $L_0/a$ with $n = 4$. In both plots, the value of $q_a$ is varied from $2$ to $8$, and $q_0 = 2$ is fixed. Computations are carried out with the side-band theory.}
 \label{fig:kxkysidebands}
\end{figure*}

Eqs. (\ref{Eq:Closure+}) and (\ref{Eq:Closure-}), reveal the effect of physical parameters, such as $L_0/a$, $\nu_0$, $m_i/m_e$, and $\rho_{*}^{-1}/a$, on the poloidal mode structure and relate them to the numerical results presented in Section \ref{Sec:numerciallinearres}. As it can be seen from the definition of $K_m$ in Eq. (\ref{eq:definitioncouplingparameters}), resistivity and finite mass-ratio effects induce a larger mode coupling and localise the mode on the low-field side. A similar effect is obtained by increasing $a/L_0$, since $G_l^{-}$ and $G_l^{+}$ are proportional to equilibrium gradients. Mode coupling is enhanced also by larger $n$ and $q$ since in Eqs. (\ref{Eq:Closure+}) and (\ref{Eq:Closure-}) the numerator scales with $k_y$ and the denominator with $1/q$. Finally, at constant aspect ratio $R_0/a$ fixed, the side-band modes $\phi_{m_0 \pm 1}$ are proportional to $1/a$, meaning that larger size is related to less mode coupling.

Following previous works \citep{rogers_dorland_2005, paolo_rogers_brunner_2008}, we reduce the Eq. (\ref{eq:modedescription}) to a Schr\"{o}dinger equation for the harmonic oscillator to obtain a dispersion relation for $\sigma$. First, as detailed in  Appendix \ref{Appendix:explicitform}, substituting $\phi_{m_0 \pm 1}$ in Eq. (\ref{eq:modedescription}) for $m_0$, and Taylor expanding around $\hat{r}$, we get
\begin{align}\label{Eq:Hermite}
    a_0 \phi_{m_0}^{''} + (b_0 + b_1 x) \phi_{m_0}^{'} + (c_0 + c_1 x + c_2 x^2)\phi_{m_0} = 0, 
\end{align} 
where $x = r - \hat{r}$,  $a_0, b_0, b_1, c_0, c_1$, and $c_2$ are constant complex coefficients.  
Taking as ansatz,  
\begin{align}
    & \phi_{m_0}  = u(x) e^{-\alpha x - \frac{\chi}{2} x^2}, \\
    & \alpha = \frac{1}{2 a_0 b_0} + \frac{\sqrt[4]{b_1^2 - 4 a_0 c_2}(\sqrt{a_0}b_0 b_1 - 2 a_0^{3/2} c_1)}{ 2 (\sqrt{a_0}(b_1^2 - 4 a_0 c_2))^{3/4}},\\   
    & \chi = \frac{b_1}{2 a_0} + \frac{1}{2}\sqrt{\frac{b_1^2 - 4 a_0 c_2}{a_0^2}},
\end{align}
and applying the coordinate transformation,
\begin{align}
    \xi = \frac{b_0 b_1 - 2 a_0 c_1}{\sqrt{2}\sqrt{a_0}(b_1^2 - 4 a_0 c_2)^{3/4}} + \frac{1}{\sqrt{2}}\Bigg(\frac{b_1^2 - 4 a_0 c_2}{a_0^2}\Bigg)^{1/4}x,
\end{align}
we obtain the Hermite equation for $u$: $\phi(\xi)  =  u(\xi) e^{-\frac{\xi^2}{2}},$ and $u''  - 2\xi u' = - 2 \lambda u$, with

\begin{align}\label{eq:lambdadef}
     & \lambda  = \frac{1}{2(b_1^2 - 4 a_0 c_2)^{3/2}}\bigg[- b_1^3 + 2 b_1^2 c_0 - 2 b_0 b_1 c_1 + 2 a_0 c_1 ^2 \notag \\
    &  + 2 b_0^2 c_2 + 4 a_0 b_1 c_2 - 8 a_0 c_0 c_2 + \sqrt{b_1^2 - 4 a_0 c_2}(4 a_0 c_2- b_1^2) \bigg]. 
\end{align}
The quantisation condition $\lambda = \delta \in \mathbb{N}$ provides the dispersion relation.

\begin{figure*}
 \centering
 \includegraphics[scale=0.35]{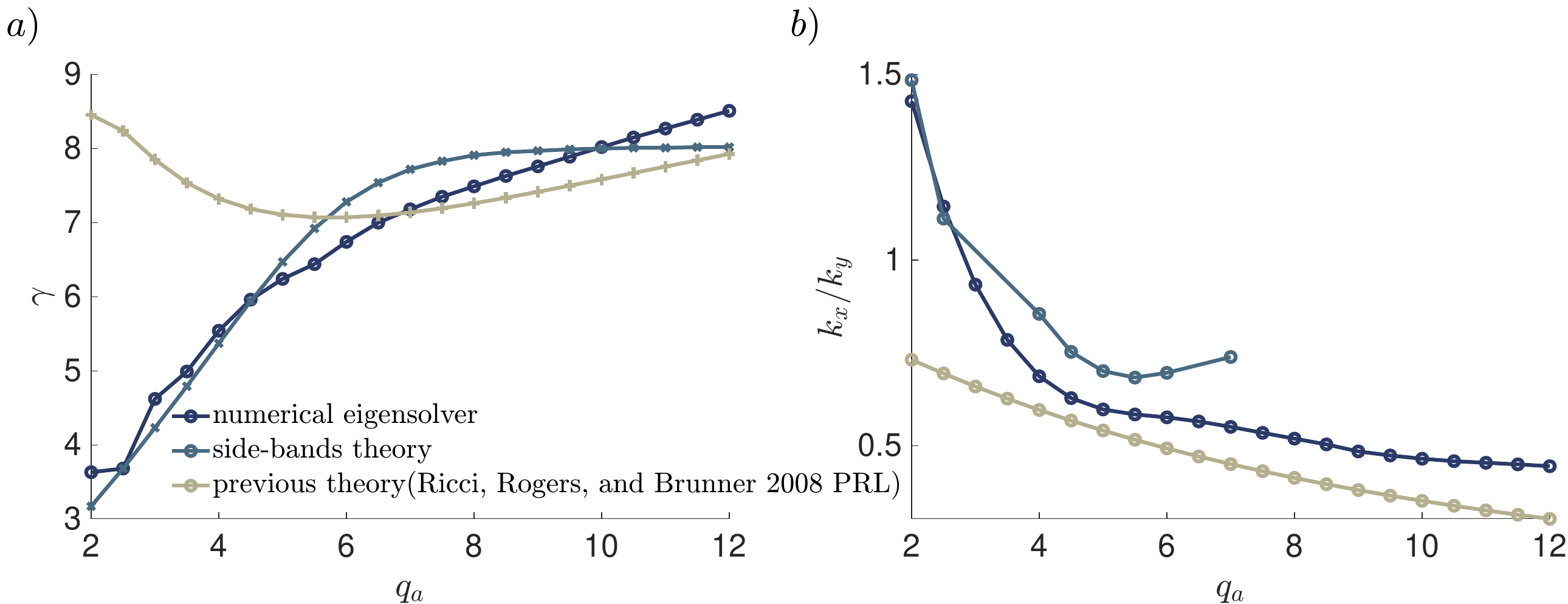}
 \caption{In $a)$, comparison between the resulting $\gamma$ from the numerical eigensolver, the side-band theory, and from the theory used in \citet{paolo_rogers_brunner_2008} with increasing $q_a$ fixing $q_0 = 2$ and $n = 3$ with $L_0 = 0.2a$. In $b)$, a similar comparison for $k_x/k_y$.}
 \label{fig:model_comp}
\end{figure*}

We solve numerically Eq. \ref{eq:lambdadef} for $\sigma$ using a Newton method, considering the solution $|\phi(\xi)| \to 0$ for $\xi \to \infty$ with eigenvalue with largest growth rate. In agreement with results in Section \ref{Sec:numerciallinearres}, it is found that the eigenvalue of the fastest growing mode is real. Moreover, the growth rate is maximised for $\delta = 0$, corresponding to an eigenmode with a Gaussian radial profile, consistent with the shape of radial profiles previously assumed in literature \citep{paolo_rogers_brunner_2008}.

The comparison of the growth rates computed with the side-band theory and with the numerical ones obtained with the eigensolver in Section \ref{Sec:numerciallinearres} is shown in Figure \ref{fig:solvervsanalytics}. A good overall agreement is obtained for varying toroidal mode number, $q$ profile, and equilibrium gradient. Only a small discrepancy between the side-band theory and the numerical eigensolver is observed for increasing values of $n$ and $q_a$. This effect is associated to non-vanishing high side-bands with $m = m_0 \pm 2, m_0 \pm 3$ ..., making the approximation in Eqs. (\ref{Eq:Closure+}) and (\ref{Eq:Closure-}) invalid.

The side-band theory allows us to reconstruct the radial and poloidal mode structure. Computing $\phi_{m_0 \pm 1}$ with Eqs. (\ref{Eq:Closure+} \ref{Eq:Closure-}) from $\phi_{m_0}$, the eigenmode is given as $\Phi(r, \theta) = \phi_{m_0 + 1}(r) e^{i(m_0 + 1)\theta} + \phi_{m_0}(r) e^{i m_0\theta} +\phi_{m_0 - 1}(r)e^{i(m_0 - 1)\theta}$. Similarly to the results in Section \ref{Sec:numerciallinearres}, $k_x \ll k_y$ for large values of shear, $n$, and $L_0$, as shown in  Figure \ref{fig:kxkysidebands}, and $k_y \lesssim k_x$ in the opposite regime. We observe that for large $n$, $q_a$, and $L_0$ values, the eigenmodes becomes strongly localised radially around $\hat{r}$, overestimating $k_x$ with respect to the results in Section \ref{Sec:numerciallinearres}. While $k_y$ is correctly estimated, a more precise computation of $k_x$ would require a larger number of side-band modes.

In conclusion, the local side-band theory provides a valid alternative to the numerical eigensolver for computing $\gamma$ and for studying the mode structure, by simplifying the eigenvalue problem in Eqs. (\ref{Eq:Eign}-\ref{Eq:EigPhi}) to the computation of the zeros of an analytical dispersion relation. While this theory correctly reconstructs the low shear poloidal structure (see Figure \ref{fig:struct2D} $a$), in the high shear case (see Figure \ref{fig:struct2D} $d$), the poloidal mode structure is poorly reproduced given the assumption of considering just three main modes.


\section{Discussion and conclusion}\label{Sec:Discussion}

This work investigates the effects of magnetic shear on ballooning-driven edge turbulence, focusing on its mode structure. Our investigation is motivated by the use of the assumption of fluctuation scale separation $1/L_p \ll k_x \ll k_y$ \citep{paolo_rogers_brunner_2008} in determining nonlinear properties of edge plasma turbulence \citep{giacomin2022turbulent}, while recent global turbulent stellarator simulations show $k_x \sim k_y$ \citep{GBS_stellarators}.

Nonlinear simulations with an idealized circular-flux surface magnetic field reveal that BM turbulence is strongly affected by the magnetic shear. At low-shear, radial and poloidal eddy sizes are comparable in magnitude, $k_x \sim k_y$, and their associated spatial scale is comparable to $L_p$, with coherent modes dominating the dynamics. In contrast, high-shear simulations recover the $k_x \ll k_y$ condition, with broad-band turbulent spectra generated by strong interaction between poloidal modes, identified as mode coupling. In both cases, turbulence shows field-aligned turbulent structures, as expected from BM turbulence.

To understand the effect of magnetic shear, we introduce a global two-dimensional linear theory for ballooning instabilities, valid at low- and high-magnetic shear, (\ref{Eq:Eign}-\ref{Eq:EigPhi}). Using the ballooning representation, Eqs. (\ref{eq:highN&T}-\ref{eq:highPhi}), we compute solutions for non-vanishing shear cases. Instead, a single-mode representation is employed for vanishing-shear cases, where the ballooning representation becomes inconsistent, Eqs. (\ref{Eq:EigZeron}-\ref{Eq:EigZeroPhi}). 

Numerical results from the linear theory qualitatively explain the mode structures observed in nonlinear simulations, showing a transition from coherent modes to a filamentary structure with a broader poloidal spectra, as the value of shear increases. We find that the poloidal mode structure is characterised by a high-frequency oscillating poloidal envelope with filamentary eddies and with dominant mode $\phi_m$ with $m \simeq n q$, thus minimizing $k_\parallel$. In the low-shear case, this relation is satisfied along the radial direction by $m = q_0 n$, where $q_0$ is the constant value of the $q$-profile, giving rise to a coherent mode with comparable radial and poloidal extensions. On the other hand, in the presence of a non-vanishing shear, the value of $m$ that minimises $k_\parallel$ depends on the radial position to maintain the field-aligned feature, giving rise to a large poloidal spectrum and providing a filamentary structure. Numerical results show that localisation on the low-field side is induced by $n$, $\hat{s}$, $L_0$, and $\nu_0$, while coherency is recovered for increasing $R_0$, at constant aspect ratio. The mode radial scale is set by the radial pressure gradient length, $L_0$, and increases with $\nu_0$. Our study confirms that the property $k_x \ll k_y$ scaling emerges only at sufficiently high shear and toroidal mode numbers, while the property $k_x \sim k_y \sim 1/L_p$ emerges in the low-shear regime.

A side-band approach that models the BM by considering interactions between a dominant poloidal mode $m_0 = n \hat{q}$, where $\hat{q}$ is the value of the $q$-profile at the rational surface, and its nearest neighbour allows us to investigate the physical mechanism influencing the mode structure transition with increasing shear. This approach shows that curvature induces mode coupling and broadens the poloidal spectrum, while finite $k_\parallel$ effects contribute to concentrate the spectrum around $m_0$. This theory efficiently computes $\gamma$, $k_x$, and $k_y$, correctly capturing the transition in mode structure between low- and high- shear. Limitations are found for large values of shear and toroidal mode numbers, corresponding to modes with large poloidal spectra, where three modes are not sufficient to completely describe the poloidal structure.

To emphasize the importance of a global linear theory, the global approaches introduced in this work are compared to the previous results in \cite{rogers2005noncurvature, paolo_rogers_brunner_2008} where the relation $k_x \sim \sqrt{k_y/L_p}$ is derived assuming that $k_y L_p \gg 1$, therefore radially studying the BM at the LFS. More details about this poloidal expansion can be found in Appendix \ref{Sec:comparisonprevtheories}.

In Figure \ref{fig:model_comp}, we compare $\gamma$, and $k_x/k_y$ from the numerical eigensolver, the side-band theory, and the poloidal expansion approach. The latter does not recover the results from the global approaches for low-shear values, while for increasing shear values, its trend is captured. This deficiency resides in the assumption of expanding around the low field side, therefore assuming the localisation of the mode in that region. However, this assumption is incorrect in the low-shear case. On the other hand, the side-band approach allows us to correctly capture $\gamma$ for both low and large shear values, and $k_x/k_y$ from low- to intermediate-shear values. However, describing the poloidal structure of high shear mode requires a more significant number of poloidal modes. Therefore, the global feature of the linear models proposed in our work avoids the poloidal \citep{paolo_rogers_brunner_2008} and radial \citep{annamaria_linear_theory} localization assumptions of previous studies, providing a robust framework to describe the properties of global BM depending on the magnetic shear.   

The work presented here offers valuable insights for the analysis of turbulence in the plasma boundary of fusion devices. For instance, it allows extending and improving nonlinear estimates for the equilibrium pressure gradient
length, the SOL width, and turbulent particle or heat fluxes \citep{giacomin2022turbulent, ricci2023theoretical, lim2024predictive}. The natural extensions of the proposed global linear theory for the BM include investigating the electrostatic BM with shaping effects \citep{riva2017plasma, lim2023effect}, in diverted configurations, with velocity and $E \times B$ shear effects, while incorporating electromagnetic contributions. This framework can also be extended to investigate DWs and KH instabilities. Despite the increased complexity, this approach can also be applied to study large-scale global modes recently identified in low-shear stellarator simulations \citep{letter_stellarators, TJK_mine, GBS_stellarators}.

\section*{Acknowledgements}
Many useful discussions with D. Mancini, L.N. Stenger, and S. Brunner are gratefully
acknowledged. The simulations presented herein were carried out in part at the Swiss National Supercomputing Centre (CSCS) under the project ID s1182, and in part using supercomputer resources provided under the EU-JA Broader Approach collaboration in the Computational Simulation Centre of International Fusion Energy Research Centre (IFERC-CSC). This work has been carried out within the framework of the EUROfusion Consortium, via the Euratom Research and Training Programme (Grant Agreement No 101052200 — EUROfusion) and funded by the Swiss State Secretariat for Education, Research and Innovation (SERI). Views and opinions expressed are however those of the author(s) only and do not necessarily reflect those of the European Union, the European Commission, or SERI. Neither the European Union nor the European Commission nor SERI can be held responsible for them. This work was supported in part by the Swiss National Science Foundation (10001273, BD). This research was also supported by a grant from the Simons Foundation (1013657, JL).

\section*{Data Availability Statement}
The data that support the findings of this study are available from the corresponding author
upon reasonable request.
\section*{Declaration of interest}
The authors report no conflict of interest.

\subsection*{Author contributions}
\textbf{Zeno Tecchiolli}: Conceptualization (equal); Data curation (equal); Formal analysis (lead); Investigation (equal); Methodology (lead); Software (lead); Validation (equal); Visualization (lead); Writing - original draft (lead);
\textbf{Antonio Coelho}: Conceptualization (equal); Investigation (equal); Software (equal); Writing - review \& editing (equal); \textbf{Joaquim Loizu}: Conceptualization (equal); Supervision (equal); Writing - review \& editing (equal); \textbf{Brenno De Lucca}: Conceptualization (equal); Validation (equal); Writing - review \& editing (equal); \textbf{Paolo Ricci}: Supervision (equal); Writing - review \& editing (equal);
\appendix
\section{Numerical implementation of the linear solvers} \label{Sec:numimplementationEig}
We use a finite difference scheme to solve all the eigenvalue problems considered in the present work. In the following, we indicate with the coordinate $y$ corresponding to the poloidal direction, $\theta$ for the solution of Eqs. (\ref{Eq:EigV}-\ref{Eq:EigPhi}) and Eqs. (\ref{Eq:EigZeron}-\ref{Eq:EigZeroPhi}), or the ballooning one $\eta$ for the solution of Eqs. (\ref{eq:highN&T}-\ref{eq:highPhi}). In the ballooning representation, $\eta \in [-\eta_{M}, \eta_{M}]$.
To avoid the coordinate singularity at $r = 0$, we set as domain $x \in (\epsilon, 1]$, with $\epsilon = 0.01/a$, and $x = r/a$. We construct a 2D grid $\textbf{x}_{ij}$ as the set of points $\textbf{x}_{ij} = (x_i, y_j)$ with $i \in [0, N_x]$,  $j \in [0, N_y]$, $x_i = (a - \epsilon) i/N_x + \epsilon$, and $y_j = 2j/N_y - 1$. The size of $\textbf{x}_{ij}$ is $N_x \times N_y$. The resulting grid is equispaced in both directions with $\Delta x = 1/N_x$ and $\Delta y = 2/N_y$. We evaluate each field on the grid points and indicate them as $N_{ij}$, $T_{ij}$, $V_{ij}$, $\Phi_{ij}$.. These matrices are vectorised to construct the vector $\boldsymbol{\Psi} = (N_{00}, ..., N_{0 j}, N_{10}, ..., N_{1j}, ..., N_{N_xN_y}, T_{00}, ..., T_{N_xN_y}, V_{00}, ..., V_{N_xN_y}, \newline \Phi_{00}, ..., \Phi_{N_xN_y})$. Consequently, the size of $\boldsymbol{\Psi}$ is $4\times N_x \times N_y$. Derivatives are discretised by using a 6th order accurate finite difference scheme.
The eigenvalue problems are recast into a matricial form as: $\sigma L\boldsymbol{\Psi} = M\boldsymbol{\Psi}$, where $L$ and $M$ are the discretisation of the left-hand-side and of the right-hand-side of the eigenvalue problems, respectively. For problems in Eqs. (\ref{Eq:EigV}-\ref{Eq:EigPhi}) and Eqs. (\ref{Eq:EigZeron}-\ref{Eq:EigZeroPhi}), we impose periodic boundary conditions in the $y$ direction by stating that $y_{i, N_y + 1} = y_{i, 0}$. Radially, we impose either homogeneous Neumann boundary conditions or homogeneous Dirichlet. No boundary condition is applied at $x = 0$ because of radial symmetry. Solving with the ballooning formalism, we set homogeneous Dirichlet at the boundary in $\eta$ to enforce the localisation of the eigenmode.

The boundary conditions effectively reduce the number of free parameters, i.e. independent elements in the vector $\boldsymbol{\Psi}$. This is typically done by imposing that a specific number of lines in the operators vanish. 
This procedure introduces a degeneracy since $\det M = 0$.  To avoid this problem, we construct a projector operator, $P$, whose action transforms the physical eigenvector, $\boldsymbol{\eta}$, which is constituted only by physical degrees of freedom, into the eigenvector with the particular boundary condition,  $\boldsymbol{\Psi} = P \boldsymbol{\eta}$. $P$ is constructed element-wise from the desired boundary conditions. Hence, the number of columns of $P$ coincides with the physical degrees of freedom, while its number of lines accounts for the number of points in the grid. Therefore, P is a rectangular matrix, while $P^{-1}$ is defined as the inverse of $P$ such that $P^{-1}P = \mathit{I}_d$. 
The boundary value problem is given by  $\gamma L \boldsymbol{\Psi}  = \gamma L P \boldsymbol{\eta} = M P \boldsymbol{\eta}$ whose inversion is $\gamma \boldsymbol{\eta} = (P^{-1}L P)^{-1} P^{-1} M P \boldsymbol{\eta}$, yielding an eigenvalue problem for the operator $(P^{-1}L P)^{-1} P^{-1} M P$, which acts in the space of physical degrees of freedom and is not degenerate. The left multiplication by $P^{-1}$ is necessary for having $P^{-1} L P$ squared and invertible. $P$ is a pseudo-projector
since the dimensions of the two rectangular matrices do not allow the product $P P$. We numerically solve the eigenvalue problem by finding the set of $\sigma$ and $\boldsymbol{\eta}$, and we obtain the constrained eigenmode via $\boldsymbol{\Psi} = P \boldsymbol{\eta}$. 
Among the values of $\sigma$ and the corresponding eigenvectors $\boldsymbol{\eta}$, the one expected to dominate the dynamics and to saturate the equilibrium  is associated with the fastest growing mode, i.e. the one with the largest real part of $\sigma$. In addition, $M$ and $L$ matrices are large-scale sparse matrices without any particular symmetry.

We solve the problem using the Arnoldi iteration  \citep{lehoucq1995analysis, stewart2002krylov}, which considers $M^\alpha \textbf{b}$ as the closest approximation to the eigenvector of a matrix $M$ corresponding to the largest $\gamma$, with starting vector $\textbf{b}$, and with a positive integer $\alpha$. This method can be tailored for finding the eigenvalue with the largest real part. The matrices are handled with the ARPACK library \cite{lehoucq1998arpack}, and the algorithm is implemented in the \texttt{eigs} function in \texttt{MATLAB} \citep{MATLAB:2024a}. We fix the $\alpha = 350$ with $10^{-8}$ as tolerance in the numerical implementation.

For each $\sigma$ found, we check that it satisfies Gerschgorin’s Circle Theorem  \citep{weisstein2003gershgorin}. We then study the influence of the grid size $N_G = N_x \times N_y$ on $\sigma$ by varying $N_x$ and $N_y$. We consider $\sigma$ converged if $\sigma$ obtained with $N_G$ grid and for $\sigma^{'}$ obtained with $N_G'$, with $N_G'>N_G$, are such that $|\gamma - \gamma'|< 10^{-4}$. Using the ballooning representation, we checked that the result is unaffected by increasing $\eta_{M}$.

\section{Derivation of Eq. (\ref{Eq:Hermite})}\label{Appendix:explicitform}
Substituting $\phi_{m_0 \pm 1}$ in Eq. (\ref{eq:modedescription}) for $m_0$, we obtain the boundary value problem:
\begin{align} \label{Eq:FormalProblem}
    K_2(r)\phi_{m_0}^{''} + K_1(r)\phi^{'}_{m_0} + K_0(r)\phi_{m_0} = 0, 
\end{align}
where $K_0$, $K_1$, and $K_2$ expressions are 
\begin{align}
    & K_0 = \frac{F_{m_0 - 1}}{K_{m_0-1}^2}\Big(G^{+}_{m_0 - 1}K^{'}_{m_0 - 1} - G^{+ '}_{m_0 - 1} K_{m_0 - 1}\Big) - \frac{G^{-}_{m_0} G^{+}_{m_0 - 1}}{K_{m_0 - 1}} \notag \\  
    & ~ ~ ~ ~ - K_{m_0} - \frac{1}{K_{m_0 + 1}^2}\Big[F_{m_0 + 1} G^{-}_{m_0 + 1} K^{'}_{m_0 + 1}   \notag \\  
    & ~ ~ ~ ~  - K_{m_0 + 1}\big(G^{-}_{m_0 + 1} G^{+}_{m_0} + F_{m_0 + 1} G^{- '}_{m_0 +1}\big)\Big], \\
    & K_1 = - \frac{1}{K_{m_0 + 1}^2}\Big\{\frac{K_{m_0 + 1}}{r K_{m_0 - 1}^2}\Big[K_{m_0 + 1}K_{m_0 - 1}\Big(- K_{m_0 - }1 \notag \\
    & ~ ~ ~ ~ + r F_{m_0 - 1}(G^{+}_{m_0 - 1} + F_{m_0}^{'})\Big)\Big]  \notag \\
    & ~ ~ ~ ~ + r F_{m_0}\Big[G_{m_0}^{+} K_{m_0 - 1}^2 +   K_{m_0 + 1}\Big(G^{-}_{m_0} K_{m_0 - 1} - F_{m_0 - 1} K^{'}_{m_0 - 1}\Big)\Big] \notag \\
    & ~ ~ ~ ~ + F_{m_0 + 1}\Big[K_{m_0 + 1}(G^{-}_{m_0 + 1} +F^{'}_{m_0}) - F_{m_0}K_{m_0 + 1}^{'}\Big]\Big\},  \\
    & K_2 = F_{m_0} \Big(-\frac{F_{m_0 - 1}}{K_{m_0 -1}}-\frac{F_{m_0 + 1}}{K_{m_0 +1}}\Big) + 1.
\end{align}
We expand $K_0, K_1,$ and $K_2$ at $r = \hat{r}$ where the equilibrium gradient peaks. Defining $x = r - \hat{r}$, we find $K_0 = c_0 + c_1 x + c_2 x^2$, $K_1 =  b_0 + b_1 x $, and $K_2 = a_0$, with $a_0, b_0, b_1, c_0, c_1, c_2$ depending on $n, q, s$ and $\sigma$, obtaining Eq. (\ref{Eq:Hermite}).

\section{Local poloidal theory for the BM}\label{Sec:comparisonprevtheories}

The system of Eqs. (\ref{Eq:Eign}-\ref{Eq:EigPhi}) can be reduced to a boundary value problem for $\Phi$ vanishing at $r = 0$ and $a$. 
Expanding the poloidal operators around $\theta = \xi = 0$ and decomposing in Fourier modes we obtain:
\begin{align}\label{Eq:Local1}
    & \phi_m^{''} + \Bigg(\frac{1}{r} +\frac{4 \rho \xi}{\sigma^2}\frac{im}{r} f_0^{'}\Bigg)\phi_m^{'}  \notag  \\
     +&\Bigg[ -\frac{m^2}{r^2} +  \frac{4 \rho}{\sigma^2}\Bigg(i m \xi \bigg( \frac{1}{r} f_0^{''}
     -\frac{1}{r^2}f_0^{'}\bigg)-  \bigg(1 -\frac{\xi^2}{2}\bigg)\frac{ m^2}{r^2}f_0^{'}\Bigg) \notag \\
    & ~ ~ ~ ~ -\frac{\frac{m_i}{m_e}\bigg(\frac{m}{q} - n\bigg)^2}{\sigma\big(\sigma + \frac{m_i}{m_e}\frac{\nu_0}{\sqrt{f_0}}\big)}\Bigg]\phi_m = 0.
\end{align}
Since Eq. (\ref{Eq:Local1}) is formally equivalent to Eq. (\ref{Eq:FormalProblem}), it is solved analogously by reducing it to a Schrödinger-like equation using an expansion around $\hat{r}$, providing a dispersion relation that we solve numerically. Given $n$, we then maximise $\gamma$ as a function of $m$ and similarly for $\xi$. We find that $\gamma$ is maximised for $\xi = 0$, $m = n \hat{q}$, and $\delta = 0$ while $\omega$ is found to vanish in these cases.

\bibliography{aipsamp}

\end{document}